\documentclass[superscriptaddress,
preprint,
nofootinbib,
nobibnotes,
 prd
]{revtex4}
\usepackage{amsfonts}
\usepackage{amsmath}
\usepackage{graphicx}
\usepackage{dcolumn}
\usepackage{bm}

\newcommand{\qed}{\nobreak \ifvmode \relax \else
      \ifdim\lastskip<1.5em \hskip-\lastskip
      \hskip1.5em plus0em minus0.5em \fi \nobreak
      \vrule height0.75em width0.5em depth0.25em\fi}

\begin{document}

\preprint{}
\title{On the (In)Efficiency of the Cross-Correlation Statistic for
Gravitational Wave Stochastic Background Signals with Non-Gaussian Noise and
Heterogeneous Detector Sensitivities}
\author{Lionel Martellini}
\email{lionel.martellini@edhec-risk.com}
\affiliation{EDHEC-Risk Institute, 400 Promenade des Anglais, BP 3116, 06202 Nice Cedex
3, France}
\affiliation{Laboratoire Artemis, Universit\'{e} C\^{o}te d'Azur, Observatoire C\^{o}te
d'Azur, CNRS, Bd de l'Observatoire CS 34229 F-06304 NICE, France}
\author{Tania Regimbau}
\affiliation{Laboratoire Artemis, Universit\'{e} C\^{o}te d'Azur, Observatoire C\^{o}te
d'Azur, CNRS, Bd de l'Observatoire CS 34229 F-06304 NICE, France}
\date{\today }

\begin{abstract}
Under standard assumptions including stationary and serially uncorrelated
Gaussian gravitational wave stochastic background signal and noise
distributions, as well as homogenous detector sensitivities, the standard
cross-correlation detection statistic is known to be optimal in the sense of
minimizing the probability of a false dismissal at a fixed value of the
probability of a false alarm. The focus of this paper is to analyze the
comparative efficiency of this statistic, versus a simple alternative
statistic obtained by cross-correlating the \textit{squared} measurements,
in situations that deviate from such standard assumptions. We find that
differences in detector sensitivities have a large impact on the comparative
efficiency of the cross-correlation detection statistic, which is dominated
by the alternative statistic when these differences reach one order of
magnitude. This effect holds even when both the signal and noise
distributions are Gaussian. While the presence of non-Gaussian signals has
no material impact for reasonable parameter values, the relative
inefficiency of the cross-correlation statistic is less prominent for
fat-tailed noise distributions but it is magnified in case noise
distributions have skewness parameters of opposite signs. Our results
suggest that introducing an alternative detection statistic can lead to
noticeable sensitivity gains when noise distributions are possibly
non-Gaussian and/or when detector sensitivities exhibit substantial
differences, a situation that is expected to hold in joint
detections from Advanced LIGO and Advanced Virgo, in particular in the early
phases of development of the detectors, or in joint detections from Advanced
LIGO and Einstein Telescope.
\end{abstract}

\pacs{Valid PACS appear here}
\maketitle





\section{\label{sec:intro} Introduction}

A stochastic background of gravitational-waves is expected to arise from the
superposition of independent signals at different stages of the evolution of
the Universe, that are too weak or too numerous to be resolved individually.
This background can be of cosmological origin, from the amplification of
vacuum fluctuations during inflation \cite%
{1975JETP...40..409G,1993PhRvD..48.3513G,1979JETPL..30..682S}, pre Big Bang
models \cite{1993APh.....1..317G,1997PhRvD..55.3330B,2010PhRvD..82h3518D},
cosmic (super)strings \cite%
{2005PhRvD..71f3510D,2007PhRvL..98k1101S,2010PhRvD..81j4028O,2012PhRvD..85f6001R}
or phase transitions \cite%
{2008PhRvD..77l4015C,2009PhRvD..79h3519C,2009JCAP...12..024C}, or of
astrophysical origin, from sources since the beginning of stellar activity
such as core collapses to neutron stars or black holes \cite%
{2005PhRvD..72h4001B,2006PhRvD..73j4024S,2009MNRAS.398..293M,2010MNRAS.409L.132Z}%
, rotating neutron stars \cite{2001A&A...376..381R,2012PhRvD..86j4007R}
including magnetars \cite%
{2006A&A...447....1R,2011MNRAS.410.2123H,2011MNRAS.411.2549M,2013PhRvD..87d2002W}%
, phase transition \cite{2009GReGr..41.1389D} or initial instabilities in
young neutron stars \cite%
{1999MNRAS.303..258F,2011ApJ...729...59Z,2004MNRAS.351.1237H,2011ApJ...729...59Z}
or compact binary mergers \cite%
{2011ApJ...739...86Z,2011PhRvD..84h4004R,2011PhRvD..84l4037M,2012PhRvD..85j4024W,2013MNRAS.431..882Z}%
.

The detection of gravitational waves has become a question of central
importance in astrophysics and the detection of the cosmological
contribution would have a profound impact on our understanding of the
evolution of the Universe, as it represents a unique window on the very
early stages up to a fraction of second after the Big Bang. An increasing
range of efforts are dedicated to the design of improved detectors. The next
generation of instruments (Advanced LIGO and Advanced Virgo \cite%
{AdLIGO,AdVIRGO}), which will start operating in 2015 and 2016 respectively,
are expected to be more than ten times more sensitive than their first
generation counterparts. Besides, third-generation interferometers such as
the European project Einstein Telescope (ET) \cite{ET}, currently under
design study, are expected to further increase the likelihood of detecting
the exceedingly small effects of gravitational waves. In parallel to the
technological efforts towards the generation of sensitivity improvements for
gravitational wave detectors, an increasing body of research is attempting
to improve upon the efficiency of data analysis methodologies involved in
stochastic gravitational wave background (SGWB) signals.

The commonly used approach to the detection of GW stochastic background
signals consists of cross-correlating the coherent measurements obtained
from a pair of detectors. Under standard assumptions including stationary
and serially uncorrelated Gaussian gravitational wave stochastic background
signal and noise distributions as well as homogenous detector sensitivities,
the cross-correlation (CC) detection statistic is known to be optimal in the
sense of minimizing the false dismissal probability at a fixed value of the
false alarm probability (see for example \cite%
{1992PhRvD..46.5250C,1993PhRvD..48.2389F,1999PhRvD..59j2001A}). Recent
predictions based on population modeling however suggest that, for many
realistic astrophysical models, there may not be enough overlapping sources,
resulting in the formation of a non-Gaussian background. It has also been
shown that the background from cosmic strings could be dominated by a
non-Gaussian contribution arising from the closest sources \cite%
{2005PhRvD..71f3510D,2012PhRvD..85f6001R}. In the past decade a few methods
have been proposed to search for a non-Gaussian stochastic background,
including the probability horizon concept developed by \cite%
{2005MNRAS.361..362C} based on the temporal evolution of the loudest
detected event on a single detector, the maximum likelihood statistic of 
\cite{2003PhRvD..67h2003D} or \cite{2014PhRvD..89l4009M}, which extends the
standard analysis in the time domain in the case of parametric or
non-parametric deviations of normality, the fourth-order correlation method
from \cite{2009PhRvD..80d3003S}, which uses fourth-order correlation between
four detectors to measure the third and the fourth moments of the
distribution of the GW signal, or the recent extension of the standard
cross-correlation statistic by \cite{2013PhRvD..87d3009T}. While most of
these papers maintain the assumption of Gaussian noise distributions so as
to better focus on the impact of deviations from normality of the signal
distribution, there is also ample evidence of strong deviations from the
Gaussian assumption for noise distributions in gravitational waves detectors
(see \cite{2002PhRvD..65l2002A,2003PhRvD..67l2002A}), and relatively little
is known about the impact of the presence of such non-Gaussian noise
distributions on the efficiency of standard methods used for the detection
of SGWB signals. Besides, the standard assumption that the two
detectors have the same sensitivity may not hold strictly for joint
observations by Advanced LIGO and Advanced Virgo \cite{AdLIGO,AdVIRGO},
especially during the early stages of development of the detectors, or joint
observations by Advanced LIGO and ET.

The focus of this paper is to analyze the efficiency of the standard CC
statistic in situations that deviate from the aforementioned standard
assumptions, and in particular involve deviations from the Gaussian
assumption and/or the presence of detectors with heterogenous sensitivities.
To do so we first introduce a simple alternative statistic obtained by
cross-correlating the \textit{squared} measurements, and we derive
closed-form expressions for the mean and variance of this statistic as a
function of the first four cumulants of the signal and noise distributions.
We also show how to obtain consistent estimates for these parameters using a
suitable extension of the likelihood function, for which we obtain an
analytical expression. These results extend our previous results 
\cite{2014PhRvD..89l4009M}, where we have focussed on a situation involving
a non-Gaussian signal distribution, but have maintained the assumption of a
Gaussian noise distribution. Turning to a numerical analysis, we find that
differences in detector sensitivities have a large impact on the comparative
efficiency of the CC detection statistic, which is dominated by the
alternative statistic when these differences reach one order of magnitude.
Remarkably, this effect holds even when both the signal and noise
distributions are Gaussian. While the presence of non-Gaussian signals has
no material impact for reasonable parameter values, we find that the
relative inefficiency of the CC statistic is less prominent in the presence
of fat-tailed noise distributions, which imply an increase in the variance
of the alternative detection statistic. On the other hand, the relative
inefficiency of the CC statistic is magnified in case noise distributions
have skewness parameters of opposite signs, a situation that leads to a
reduction in the variance of the alternative detection statistic through a
diversification effect. Overall, our results suggest that introducing an
alternative detection statistic can potentially lead to noticeable
sensitivity gains when noise distributions are non-Gaussian and/or when
detector sensitivities exhibit substantial differences.

The rest of the paper is organized as follows. In Section 2, we provide a
brief review of the standard cross correlation statistic and introduce the
alternative detection statistic. In Section 3, we perform a comparative
analysis of the efficiency of the CC detection statistic versus the
alternative detection statistic, and show that the latter dominates the
former in a number of cases of potential practical relevance. In Section 4,
we extend the maximum likelihood estimation techniques to a situation
involving potentially non-Gaussian signal and non-Gaussian noise
distributions so as to obtain consistent estimators not only for the
variance but also the skewness and kurtosis of the signal and noise
distributions, which are needed for implementing the alternative statistic.
Finally, Section 5 contains a conclusion and suggestions for further
research.

\section{Introducing a New Detection Statistic}

In this Section, we first recall standard results related to the
cross-correlation statistic. We then introduce an alternative statistic
given by the cross-correlation of squared measurements.

\subsection{Assumptions and Notation}

Consider two gravitational wave detectors. The output of each detector is a
collection of dimensionless strain measurements. Suppose that $N$ such
measurements are made by each detector at regular time intervals. Denote
these measurements by a $T\times 2$ matrix $h$ with components $h_{t}^{k}$,
where $i=1,2$ labels the detector, and $t$ $=1,2,...,N$ is the discrete date
of measurement. To determine whether or not the data $h$ contains some
desired signal, one usually compares the value of some detection statistic $%
\Gamma \left( h\right) $ to some threshold value $\Gamma _{\ast }$. If $%
\Gamma \left( h\right) $ is greater than the threshold value $\Gamma _{\ast
} $, one concludes that a signal is present and otherwise one concludes that
no signal is present. A detection statistic is said to be optimal if it
yields the smallest probability of mistakenly concluding a signal is present
(probability of a false alarm, or pfa) after choosing a threshold which
fixes the probability for mistakenly concluding that a signal is absent
(probability of a false dismissal, or pfd).

We first decompose the measurement output for detector $i$ in terms of noise
versus signal, which gives when written in terms of random variables: 
\begin{equation*}
\mathcal{H}_{i}\mathcal{=N}_{i}\mathcal{+S}_{i} 
\end{equation*}%
where $\mathcal{N}_{i}$ denotes the noise detected by the detector $i$ and $%
\mathcal{S}_{i}$ denotes the signal detected by the detector $i$ so that $%
\mathcal{H}_{i}$ is the total measurement for the detector $i$. If we now
assume that the detectors are coincident and coaligned (i.e., they have
identical location and arm orientations), we obtain that the signal received
by both detectors is drawn from the same distribution. Under this
assumption, we would have that: 
\begin{equation*}
\mathcal{S}_{1}=\mathcal{S}_{2}\equiv \mathcal{S} 
\end{equation*}

In terms of the realization of such random variables for either one of the
two detectors, we note: 
\begin{equation*}
h_{it}=n_{it}+s_{t} 
\end{equation*}

Given that both signal and noise distributions can potentially be
non-Gaussian, we denoted by $c_{j}$, $j=1,2,3,4$, the first four \textit{%
cumulants} of the signal distribution, and by $c_{i,j}$, $j=1,2,3,4$, the
first four \textit{cumulants} of the noise distribution for detector $i$,
with $i=1$ or $2$.

Let us recall that for a random variable $X$ with density function denoted
by $f_{X}$ (here $X=S$ or $\mathcal{N}_{i}$, for $i=1,2$), we can introduce
the \textit{moment generating function}: 
\begin{equation}
M_{X}\left( t\right) =\mathbb{E}\left[ e^{tX}\right] =\int\nolimits_{-\infty
}^{\infty }e^{tx}f_{X}\left( x\right) dx
\end{equation}%
which is related to the characteristic distribution $\psi _{X}$, i.e., the
Fourier transform of the function $f_{X}$, by $\psi _{X}\left( t\right)
=M_{X}\left( it\right) $. The $j^{th}$ (non central) moment of the
distribution of the random variable $X$ is given by the $j^{th}$ derivative
of the moment-generating function $M_{X}$ taken at $t=0$ (hence the name 
\textit{moment generating function}): $\mu _{j}=M_{X}^{\left( j\right)
}\left( 0\right) =\left( -i\right) ^{j}\psi _{X}^{\left( j\right) }\left(
0\right) $. Using the Taylor expansion of the exponential function around 0, 
$e^{x}=\sum\limits_{j=0}^{\infty }\dfrac{x^{j}}{j!}$, we obtain a new
expression for the characteristic function: 
\begin{equation}
\psi _{X}\left( t\right) =\mathbb{E}\left[ e^{itX}\right] =\sum%
\limits_{j=0}^{\infty }\dfrac{\left( it\right) ^{j}}{j!}E\left( X^{j}\right)
\equiv \sum\limits_{j=0}^{\infty }\dfrac{\left( it\right) ^{j}}{j!}\mu _{j}
\end{equation}

We also introduce the \textit{cumulant generating function} $m_{X}$ as the
logarithm of the moment generating function: 
\begin{equation}
m_{X}\left( t\right) =\log M_{X}\left( t\right) =\log
\sum\limits_{j=1}^{\infty }\dfrac{\left( t\right) ^{j}}{j!}\mu _{j}
\end{equation}

A Taylor expansion of the cumulant generating function $m_{X}$ is given by a
series of the following form: 
\begin{equation}
m_{X}\left( t\right) =m_{X}\left( 0\right) +\sum\limits_{j=1}^{\infty }%
\dfrac{t^{j}}{j!}m_{X}^{\left( j\right) }\left( 0\right)
\end{equation}
and we define $c_{j}=m_{X}^{\left( j\right) }\left( 0\right) $ as the $%
j^{th} $ \textit{cumulant} of the random variable $X.$A moments-to-cumulants
relationship can be obtained by expanding the exponential and equating
coefficients of $t^{j}$ in: 
\begin{equation}
M_{X}\left( t\right) =\exp \left[ m_{X}\left( t\right) \right]
\Longleftrightarrow \sum\limits_{j=0}^{\infty }\dfrac{t^{j}}{j!}\mu
_{j}=\exp \left[ \sum\limits_{j=1}^{\infty }\dfrac{t^{j}}{j!}c_{j}\right] .
\end{equation}

Conversely, a cumulants-to-moments relationship is obtained by expanding the
logarithmic and equating coefficients of $t^{j}$ in $m_{X}\left( t\right)
=\log M_{X}\left( t\right) $. In particular we have: 
\begin{eqnarray}
c_{1} &=&m_{s}^{\prime }\left( 0\right) =\mu _{1}=\mu  \label{c1} \\
c_{2} &=&m_{s}^{\prime \prime }\left( 0\right) =\mu _{2}-\mu _{1}^{2}=\sigma
^{2}  \label{c2} \\
c_{3} &=&m_{s}^{(3)}\left( 0\right) =\mu _{3}-3\mu _{2}\mu _{1}+2\mu _{1}^{3}
\label{c3} \\
c_{4} &=&m_{s}^{(4)}\left( 0\right) =\mu _{4}-4\mu _{3}\mu _{1}-3\mu
_{2}^{2}+12\mu _{2}\mu _{1}^{2}-6\mu _{1}^{4}  \label{c4}
\end{eqnarray}

We note that the first cumulant is equal to the first moment (the mean), and
the second cumulant is equal to the second-centered moment (the variance).
For the Gaussian distribution with mean $\mu $ and variance $\sigma ^{2}$,
we have $c_{1}=\mu $, $c_{2}=\sigma ^{2}$, and $c_{k}=0$ for $k>2$. This
allows us to identify deviations from the Gaussian assumption through the
presence of non-zero 3rd- and 4th-order cumulants, $c_{3}$ and $c_{4}$,
which are sometimes normalized so as to transform into \textit{skewness} and 
\textit{kurtosis} parameters, respectively defined as: $skw=\dfrac{c_{3}}{%
c_{2}^{3/2}}$ and $kurt=\dfrac{c_{4}}{c_{2}^{2}}$. In our application, it
should be noted that signal and noise distributions are centered and
therefore we have $c_{1}=c_{1,1}=c_{2,1}=0.$ We also use the notation $%
c_{2}=\alpha ^{2}$, $c_{1,2}=\sigma _{1}^{2}$, and $c_{2,2}=\sigma _{2}^{2}$%
, where $\alpha $, $\sigma _{1}$, and $\sigma _{2}$ denote the
standard-deviations for the signal, detector 1 and detector 2 distributions,
respectively.

\subsection{Distribution of the Cross-Correlation Detection Statistic}

We first define the standard cross correlation detection statistic $DS_{cc}$
as: 
\begin{equation}
DS_{cc}=\dfrac{1}{T}\sum\limits_{t=1}^{T}\mathcal{H}_{1t}\mathcal{H}_{2t}
\end{equation}%
where $\mathcal{H}_{it}=\mathcal{N}_{it}+\mathcal{S}_{t}$, for $i=1,2$, and
where $\mathcal{N}_{it}$ and $\mathcal{S}_{t}$, for $1\leq t\leq T,$ are $T$
independent copies of the random variables $\mathcal{N}_{i}$ and $\mathcal{S}
$, respectively.

We have: 
\begin{equation}
DS_{cc}=\dfrac{1}{T}\left( \sum\limits_{t=1}^{T}\mathcal{N}_{1t}\mathcal{N}%
_{2t}+\sum\limits_{t=1}^{T}\mathcal{N}_{1t}\mathcal{S}_{t}+\sum%
\limits_{t=1}^{T}\mathcal{N}_{2t}\mathcal{S}_{t}+\sum\limits_{t=1}^{T}%
\mathcal{S}_{t}^{2}\right)
\end{equation}

A signal is presumed to be detected when the detection statistic $DS_{cc}$
exceeds a given detection threshold $DT$: 
\begin{equation}
DS_{cc}>DT
\end{equation}

We typically select the detection threshold $DT$ such that $pfa=x\%$, for a
given confidence level $x\%$, where the probability of a false alarm is
given by the probability to exceed the threshold in a situation where there
is no signal: 
\begin{equation}
pfa=\Pr \left( \left. DS_{cc}>DT\right\vert \mathcal{H}_{it}=\mathcal{N}%
_{it}\right)
\end{equation}

Obviously, $pfa$ is independent of the signal distribution. What depends on
the signal distribution is the probability of a false dismissal $pfd$ given
by the probability that the detection statistic remains below the threshold
even if there is a signal: 
\begin{equation}
pfd=\Pr \left( \left. DS_{cc}<DT\right\vert \mathcal{H}_{it}=\mathcal{N}%
_{it}+\mathcal{S}_{t}\right)
\end{equation}

By the central limit theorem, it can be shown that the detection statistic $%
DS_{cc}$ is asymptotically normally distributed whether or not the signal
and noise distributions are Gaussian (see \cite{2014PhRvD..89l4009M} for
more details in the case of a non-Gaussian signal). In this situation, the
distribution of the CC detection statistic is fully characterized by its
mean and variance, which can be explicitly obtained as follows: 
\begin{eqnarray*}
\mathbb{E}\left[ DS_{cc}\right] &=&\dfrac{1}{T}\mathbb{E}\left[
\sum\limits_{t=1}^{T}\mathcal{S}_{t}^{2}\right] =\dfrac{1}{T}T\mathbb{E}%
\left( \mathcal{S}^{2}\right) =\dfrac{1}{T}T\mathbb{V}ar\left[ \mathcal{S}%
\right] =\alpha ^{2} \\
\mathbb{V}ar\left[ DS_{cc}\right] &=&\dfrac{1}{T^{2}}\mathbb{V}ar\left[
\sum\limits_{t=1}^{T}\mathcal{N}_{1t}\mathcal{N}_{2t}+\sum\limits_{t=1}^{T}%
\mathcal{N}_{1t}\mathcal{S}_{t}+\sum\limits_{t=1}^{T}\mathcal{N}_{2t}%
\mathcal{S}_{t}+\sum\limits_{t=1}^{T}\mathcal{S}_{t}^{2}\right] \\
&=&\dfrac{1}{T^{2}}T\left[ \mathbb{V}ar\left( \mathcal{N}_{1}\mathcal{N}%
_{2}\right) +\mathbb{V}ar\left( \mathcal{N}_{1}\mathcal{S}\right) +\mathbb{V}%
ar\left( \mathcal{N}_{2}\mathcal{S}\right) +\mathbb{V}ar\left( \mathcal{S}%
^{2}\right) \right] \\
&=&\dfrac{1}{T}\left( \mathbb{V}ar\left( \mathcal{N}_{1}\right) \mathbb{V}%
ar\left( \mathcal{N}_{2}\right) +\mathbb{V}ar\left( \mathcal{N}_{1}\right) 
\mathbb{V}ar\left( \mathcal{S}\right) +\mathbb{V}ar\left( \mathcal{N}%
_{2}\right) \mathbb{V}ar\left( \mathcal{S}\right) +\mathbb{E}\left( \mathcal{%
S}^{4}\right) -\left( \mathbb{E}\left( \mathcal{S}^{2}\right) \right)
^{2}\right) \\
&=&\dfrac{1}{T}\left( \sigma _{1}^{2}\sigma _{2}^{2}+\sigma _{1}^{2}\alpha
^{2}+\sigma _{2}^{2}\alpha ^{2}+c_{4}+3\alpha ^{4}-\alpha ^{4}\right) =%
\dfrac{1}{T}\left( \sigma _{1}^{2}\sigma _{2}^{2}+\sigma _{1}^{2}\alpha
^{2}+\sigma _{2}^{2}\alpha ^{2}+2\alpha ^{4}+c_{4}\right)
\end{eqnarray*}

Hence, we obtain that for general signal and noise distributions the
cross-correlation detection statistic $DS_{cc}$ is asymptotically normally
distributed, with mean $\alpha ^{2}$ and variance given by $\dfrac{1}{T}%
\left( \sigma _{1}^{2}\sigma _{2}^{2}+\sigma _{1}^{2}\alpha ^{2}+\sigma
_{2}^{2}\alpha ^{2}+2\alpha ^{4}+c_{4}\right) $. It should be noted that the
variance of the detection statistic is identical whether or not the noise
distributions are Gaussian since it does \textit{not} depend on the higher
order cumulants of the noise distributions, while it depends on the
fourth-order cumulant of the signal distribution. This is because the
standard cross-correlation detection statistic involves the squared value of
the signal distributions, while the noise distributions are not squared. In
what follows, we discuss the introduction of a new detection statistic that
would make the detection procedure explicitly dependent upon the higher
order cumulants of the noise distribution, and which will be found to
dominate the cross correlation statistic for some realistic parameter values.

\subsection{Introducing an Alternative Detection Statistic}

For a Gaussian signal, the cross-correlation detection statistic can be
shown to be optimal in the sense of minimizing the false dismissal
probability at a fixed value of the false alarm probability, a result which
holds under restrictive assumptions \cite{2003PhRvD..67h2003D} including
stationary and serially uncorrelated Gaussian gravitational wave stochastic
background signal and noise distributions. In the general non-Gaussian case,
the cross-correlation detection statistic may not be optimal, and may be
dominated by an alternative detection statistic, which can be written in
general as \textit{some} function of the observations $f\left( \mathcal{H}%
_{1t},\mathcal{H}_{2t}\right) _{t=1,...,T}\neq \dfrac{1}{T}%
\sum\limits_{t=1}^{T}\mathcal{H}_{1t}\mathcal{H}_{2t}$. It is unclear how
one could derive an optimal detection statistic in a fully general setting,
and we introduce in what follows a simple heuristic alternative detection
statistic, denoted by $DS_{alt}$, which is given by the cross-correlation of 
\textit{squared} detector measurements:%
\begin{equation}
DS_{alt}=\dfrac{1}{T}\sum\limits_{t=1}^{T}\mathcal{H}_{1t}^{2}\mathcal{H}%
_{2t}^{2}
\end{equation}

By the central limit theorem, we know again that $DS_{alt}$ is
asymptotically Gaussian, and the asymptotic distribution for the detection
statistic is therefore fully characterized by its first two moments $\mu
_{alt}$ and $\sigma _{alt}^{2}$. We first have: 
\begin{eqnarray}
\mu _{alt} &=&\mathbb{E}\left[ DS_{alt}\right] =\dfrac{1}{T}\mathbb{E}\left[
\sum\limits_{t=1}^{T}\left( \mathcal{N}_{1t}+\mathcal{S}_{t}\right)
^{2}\left( \mathcal{N}_{2t}+\mathcal{S}_{t}\right) ^{2}\right]  \label{mualt}
\\
&=&\dfrac{1}{T}\left( \sum\limits_{t=1}^{T}\mathbb{E}\left( \mathcal{N}%
_{1t}^{2}\mathcal{N}_{2t}^{2}\right) +\mathbb{E}\left( \mathcal{S}_{t}^{2}%
\mathcal{N}_{1t}^{2}\right) +\mathbb{E}\left( \mathcal{S}_{t}^{2}\mathcal{N}%
_{2t}^{2}\right) +\mathbb{E}\left( \mathcal{S}_{t}^{4}\right) \right) \\
&=&\sigma _{1}^{2}\sigma _{2}^{2}+\sigma _{1}^{2}\alpha ^{2}+\sigma
_{2}^{2}\alpha ^{2}+3\alpha ^{4}+c_{4}
\end{eqnarray}

In case the signal is absent ($c_{4}=\alpha ^{2}=0$), the expression for the
mean value of the alternative detection statistic (denoted by $\mu
_{alt}^{ns}$ in this case, where \textit{ns} stands for \textit{no signal})
further simplifies into:%
\begin{equation}
\mu _{alt}^{ns}=\sigma _{1}^{2}\sigma _{2}^{2}  \label{mualtnosignal}
\end{equation}

We then compute $\sigma _{alt}^{2}=\mathbb{V}ar\left( DS_{alt}^{2}\right) $: 
\begin{eqnarray*}
\sigma _{alt}^{2} &=&\dfrac{1}{T^{2}}\mathbb{V}ar\left[ \sum%
\limits_{t=1}^{T}\left( \mathcal{N}_{1t}+\mathcal{S}_{t}\right) ^{2}\left( 
\mathcal{N}_{2t}+\mathcal{S}_{t}\right) ^{2}\right] \\
&=&\dfrac{1}{T^{2}}\sum\limits_{t=1}^{T}\mathbb{V}ar\left[ \left( \mathcal{N}%
_{1t}+\mathcal{S}_{t}\right) ^{2}\left( \mathcal{N}_{2t}+\mathcal{S}%
_{t}\right) ^{2}\right] \\
&=&\dfrac{1}{T}\mathbb{V}ar\left[ \left( \mathcal{N}_{1}+\mathcal{S}\right)
^{2}\left( \mathcal{N}_{2}+\mathcal{S}\right) ^{2}\right]
\end{eqnarray*}

We note that: 
\begin{equation*}
\mathbb{V}ar\left[ \left( \mathcal{N}_{1}+\mathcal{S}\right) ^{2}\left( 
\mathcal{N}_{2}+\mathcal{S}\right) ^{2}\right] =\mathbb{E}\left[ \left( 
\mathcal{N}_{1}+\mathcal{S}\right) ^{4}\left( \mathcal{N}_{2}+\mathcal{S}%
\right) ^{4}\right] -\left( \mathbb{E}\left[ \left( \mathcal{N}_{1}+\mathcal{%
S}\right) ^{2}\left( \mathcal{N}_{2}+\mathcal{S}\right) ^{2}\right] \right)
^{2}
\end{equation*}

After some tedious computations, we first obtain: 
\begin{eqnarray}
\mathbb{E}\left[ \left( \mathcal{N}_{1}+\mathcal{S}\right) ^{2}\left( 
\mathcal{N}_{2}+\mathcal{S}\right) ^{2}\right]  &=&\mu _{8}+6\mu _{6}\mu
_{1,2}+6\mu _{6}\mu _{2,2}+4\mu _{5}\mu _{1,3}+4\mu _{5}\mu _{2,3}+\mu
_{4}\mu _{1,4}+6\mu _{4}\mu _{2,4}  \notag \\
&&+36\mu _{4}\mu _{1,2}\mu _{2,2}+24\mu _{3}\mu _{1,3}\mu _{2,2}+24\mu
_{3}\mu _{1,2}\mu _{2,3}+6\mu _{2}\mu _{1,4}\mu _{2,2}+6\mu _{2}\mu
_{1,2}\mu _{2,4}  \notag \\
&&+16\mu _{2}\mu _{1,3}\mu _{2,3}+\mu _{1,4}\mu _{2,4}
\end{eqnarray}%
where we use the following notation for the higher order moments of the
signal and noise distributions for $j=1,...,8$:%
\begin{eqnarray*}
\mu _{j} &=&\mathbb{E}\left[ \mathcal{S}^{j}\right]  \\
\mu _{1,j} &=&\mathbb{E}\left[ \mathcal{N}_{1}^{j}\right]  \\
\mu _{2,j} &=&\mathbb{E}\left[ \mathcal{N}_{1}^{j}\right] 
\end{eqnarray*}

Finally we obtain:

\begin{equation*}
T\sigma _{alt}^{2}=\mathbb{E}\left[ \left( \mathcal{N}_{1}+\mathcal{S}%
\right) ^{4}\left( \mathcal{N}_{2}+\mathcal{S}\right) ^{4}\right] -\left( 
\mathbb{E}\left[ \left( \mathcal{N}_{1}+\mathcal{S}\right) ^{2}\left( 
\mathcal{N}_{2}+\mathcal{S}\right) ^{2}\right] \right) ^{2} 
\end{equation*}%
which after more tedious calculation gives the following expression for the
variance of the alternative detection statistic:%
\begin{eqnarray}
\sigma _{alt}^{2} &=&\dfrac{1}{T}(\mu _{8}-\mu _{4}^{2}+2\left( 3\mu
_{6}-\mu _{2}\mu _{4}\right) \left( \mu _{1,2}+\mu _{2,2}\right) +4\mu
_{5}\left( \mu _{1,3}+\mu _{2,3}\right)  \notag \\
&&+\mu _{4}\mu _{1,4}-\mu _{2}^{2}\mu _{1,2}^{2}+\mu _{4}\mu _{2,4}-\mu
_{2}^{2}\mu _{2,2}^{2}+2\left( 17\mu _{4}-\mu _{2}^{2}\right) \mu _{1,2}\mu
_{2,2}  \notag \\
&&+24\mu _{3}\left( \mu _{1,3}\mu _{2,2}+\mu _{1,2}\mu _{2,3}\right) +16\mu
_{2}\mu _{1,3}\mu _{2,3}+2\mu _{2}\mu _{2,2}\left( 3\mu _{1,4}-\mu
_{1,2}^{2}\right)  \notag \\
&&+2\mu _{2}\mu _{1,2}\left( 3\mu _{2,4}-\mu _{2,2}^{2}\right) +\mu
_{1,4}\mu _{2,4}-\mu _{1,2}^{2}\mu _{2,2}^{2})
\end{eqnarray}

If we now assume a symmetric signal distribution ($\mu _{1},\mu _{3},\mu
_{5},\mu _{7}=0$), the expression for the variance of the detection
statistic simplifies into:%
\begin{eqnarray}
\sigma _{alt}^{2} &=&\dfrac{1}{T}(\mu _{8}-\mu _{4}^{2}+2\left( 3\mu
_{6}-\mu _{2}\mu _{4}\right) \left( \mu _{1,2}+\mu _{2,2}\right) \\
&&+\mu _{4}\mu _{1,4}-\mu _{2}^{2}\mu _{1,2}^{2}+\mu _{4}\mu _{2,4}-\mu
_{2}^{2}\mu _{2,2}^{2}+2\left( 17\mu _{4}-\mu _{2}^{2}\right) \mu _{1,2}\mu
_{2,2}  \notag \\
&&+16\mu _{2}\mu _{1,3}\mu _{2,3}+2\mu _{2}\mu _{2,2}\left( 3\mu _{1,4}-\mu
_{1,2}^{2}\right)  \notag \\
&&+2\mu _{2}\mu _{1,2}\left( 3\mu _{2,4}-\mu _{2,2}^{2}\right) +\mu
_{1,4}\mu _{2,4}-\mu _{1,2}^{2}\mu _{2,2}^{2})  \label{varaltsym}
\end{eqnarray}

If the noise distributions are also symmetric ($\mu _{1,3},\mu _{2,3}=0$),
the expression further simplifies into:%
\begin{eqnarray}
\sigma _{alt}^{2} &=&\dfrac{1}{T}(\mu _{8}-\mu _{4}^{2}+2\left( 3\mu
_{6}-\mu _{2}\mu _{4}\right) \left( \mu _{1,2}+\mu _{2,2}\right)  \notag \\
&&+\mu _{4}\mu _{1,4}-\mu _{2}^{2}\mu _{1,2}^{2}+\mu _{4}\mu _{2,4}-\mu
_{2}^{2}\mu _{2,2}^{2}+2\left( 17\mu _{4}-\mu _{2}^{2}\right) \mu _{1,2}\mu
_{2,2}  \notag \\
&&+2\mu _{2}\mu _{2,2}\left( 3\mu _{1,4}-\mu _{1,2}^{2}\right) +2\mu _{2}\mu
_{1,2}\left( 3\mu _{2,4}-\mu _{2,2}^{2}\right) +\mu _{1,4}\mu _{2,4}-\mu
_{1,2}^{2}\mu _{2,2}^{2})  \label{varalt}
\end{eqnarray}

In case the signal is absent ($\mu _{8}=\mu _{6}=\mu _{4}=\mu _{2}=0$), the
expression for the variance of the alternative detection statistic (denoted
by $\sigma _{alt}^{2,ns}$) becomes:%
\begin{equation}
\sigma _{alt}^{2,ns}=\dfrac{1}{T}\left( \mu _{1,4}\mu _{2,4}-\mu
_{1,2}^{2}\mu _{2,2}^{2}\right) =T\left( \left( c_{1,4}+3\sigma
_{1}^{4}\right) \left( c_{2,4}+3\sigma _{2}^{4}\right) -\sigma
_{1}^{4}\sigma _{2}^{4}\right)  \label{varaltnosignal}
\end{equation}

Clearly, the variance of the alternative detection statistic is higher when
the higher order cumulants of the noise are not zero, which will have
implications for the sensitivity of the detection procedure. The higher
variance of the distribution of the alternative detection statistic in the
non-Gaussian case (that is the case when noise distribution are potentially
non-Gaussian) implies that the alternative detection statistic has fatter
tails compared the Gaussian case (that is the case when noise distributions
are Gaussian). As a result, the detection threshold corresponding to a given
pfd will be lower in the non-Gaussian case, which in turn allows for the
detection of fainter signals when the presence of non-Gaussianity is taken
in to account with respect to a situation where the observer uses the
alternative detection statistic while wrongly assuming that the underlying
signal and noise distributions are Gaussian. In other words, for an observer
using the alternative detection statistic, taking into account the
non-Gaussianity of the signal and noise distributions will improve the
detection methodology.

An outstanding question, however, remains with respect to whether or not the
observer would be better off using the alternative versus the standard
cross-correlation statistic. While the alternative detection statistic has
no claim to optimality, and is expected to be dominated by the
cross-correlation statistic when signal and noise distributions are
Gaussian, it may in principle dominate the standard cross-correlation
statistic in the non-Gaussian case since the optimality of the CC statistic
has not been established in this more general setting.

\section{Comparative Efficiency of the Cross-Correlation versus the
Alternative Detection Statistic}

To compare the performance of the standard and alternative detection
statistic in the general non-Gaussian case, we use the following multi-step
procedure.

\begin{itemize}
\item Step 1: We select a set of parameter values for the signal and noise
distributions, and we apply the transformation that allows one to turn the
cumulants into corresponding moments, which will be needed in the
expressions for the mean and variance of the alternative detection
statistic. Assuming for simplicity symmetric signal and noise distributions,
we have:%
\begin{eqnarray}
\mu _{2} &=&c_{2}=\alpha ^{2},\mu _{1,2}=c_{1,2}=\sigma _{1}^{2},\mu
_{2,2}=c_{2,2}=\sigma _{2}^{2} \\
\mu _{4} &=&c_{4}+3c_{2}^{2},\mu _{1,4}=c_{1,4}+3c_{1,2}^{2},\mu
_{2,4}=c_{2,4}+3c_{2,2}^{2} \\
\mu _{6} &=&c_{6}+15c_{4}c_{2}+15c_{2}^{3} \\
\mu _{8} &=&c_{8}+28c_{6}c_{2}+35c_{4}^{2}+210c_{4}c_{2}^{2}+105c_{2}^{4}
\end{eqnarray}%
We may then obtain the corresponding values for $\mu _{alt}$ and $\sigma
_{alt}^{2}$ in the presence of the signal, using equations (\ref{mualt}) and
(\ref{varalt}), as well as the corresponding values for $\mu _{alt}^{ns}$
and $\sigma _{alt}^{2,ns}$ in the absence of the signal, using equations (%
\ref{mualtnosignal}) and (\ref{varaltnosignal}).

\item Step 2: We select a given pfa value, taken in the numerical analysis
that follows to be 5\%, and obtain the corresponding thresholds for both the
standard and alternative detection statistics, denoted respectively by $%
DS_{cc}$ and $DS_{alt}$ (which are functions of the selected pfa value),
using the following equations: 
\begin{eqnarray}
pfa &=&\Pr \left( \left. DS_{cc}>DT_{cc}\left( pfa\right) \right\vert 
\mathcal{H}_{it}=\mathcal{N}_{it}\right)  \\
&=&\Pr \left( \left. DS_{alt}>DT_{alt}\left( pfa\right) \right\vert \mathcal{%
H}_{it}=\mathcal{N}_{it}\right) 
\end{eqnarray}%
based upon the following Gaussian distributions for the detection statistic
in case the signal is absent: 
\begin{eqnarray*}
DS_{cc} &=&\dfrac{1}{T}\sum\limits_{t=1}^{T}\mathcal{H}_{1t}\mathcal{H}_{2t}%
\underset{T\rightarrow \infty }{\sim }\mathcal{N}\left( 0,\dfrac{1}{T}\sigma
_{1}^{2}\sigma _{2}^{2}\right)  \\
DT_{alt} &=&\dfrac{1}{T}\sum\limits_{t=1}^{T}\mathcal{H}_{1t}^{2}\mathcal{H}%
_{2t}^{2}\underset{T\rightarrow \infty }{\sim }\mathcal{N}\left( \mu
_{alt}^{ns}=\sigma _{1}^{2}\sigma _{2}^{2},\sigma _{alt}^{2,ns}=\dfrac{1}{T}%
\left( \left( c_{1,4}+3\sigma _{1}^{4}\right) \left( c_{2,4}+3\sigma
_{2}^{4}\right) -\sigma _{1}^{4}\sigma _{2}^{4}\right) \right) 
\end{eqnarray*}%
where we take $T=10^{5}$ in the base case.

\item Step 3: We compute the probability of a false dismissal corresponding
to the standard cross-correlation statistic and the probability of a false
dismissal corresponding to the alternative detection statistic using: 
\begin{eqnarray}
pfd_{cc} &=&\Pr \left( \left. DS_{cc}<DT_{cc}\left( pfa\right) \right\vert 
\mathcal{H}_{it}=\mathcal{N}_{it}+\mathcal{S}_{t}\right) \\
pfd_{alt} &=&\Pr \left( \left. DS_{alt}<DT_{alt}\left( pfa\right)
\right\vert \mathcal{H}_{it}=\mathcal{N}_{it}+\mathcal{S}_{t}\right)
\end{eqnarray}

based upon the following Gaussian distributions for the detection statistic
in case the signal is present: 
\begin{eqnarray*}
DS_{cc} &=&\dfrac{1}{T}\sum\limits_{t=1}^{T}\mathcal{H}_{1t}\mathcal{H}_{2t}%
\underset{T\rightarrow \infty }{\sim }\mathcal{N}\left( \alpha ^{2},\dfrac{1%
}{T}\left( \sigma _{1}^{2}\sigma _{2}^{2}+\sigma _{1}^{2}\alpha ^{2}+\sigma
_{2}^{2}\alpha ^{2}+2\alpha ^{4}+c_{4}\right) \right)  \\
DT_{alt} &=&\dfrac{1}{T}\sum\limits_{t=1}^{T}\mathcal{H}_{1t}^{2}\mathcal{H}%
_{2t}^{2}\underset{T\rightarrow \infty }{\sim }\mathcal{N}\left( \mu
_{alt}^{ns},\sigma _{alt}^{2}\right) 
\end{eqnarray*}
\end{itemize}

If we can find a set of parameter values and a pfa value such that $%
pfd_{alt}<$ $pfd_{cc}$, we would then prove that the standard
cross-correlation analysis is not always optimal, and we would also show
that the alternative statistic we have introduced allows for a more
efficient detection procedure, at least for the selected set of parameter
values. In what follows, we analyze the probability of a false alarm for
both statistics in various situations involving homogenous versus
heterogenous detector sensitivities, as well as various assumptions
regarding the higher order moments of the noise distributions. Note that the
signal distribution is assumed to be Gaussian in all the result that we
present below. In unreported results, we have analyzed the relative
efficiency of the two statistics in situations involving a non-Gaussian
signal, and have found only very small differences with respect to the
Gaussian signal case. Indeed, the signal is assumed to be small compared to
the noise in realistic situations, and therefore the impact of deviations
from the Gaussian assumption at the signal level will be dwarfed by the
impact of deviations from the Gaussian assumption at the noise level.

\begin{figure}[tbp]
\includegraphics[width=\textwidth]{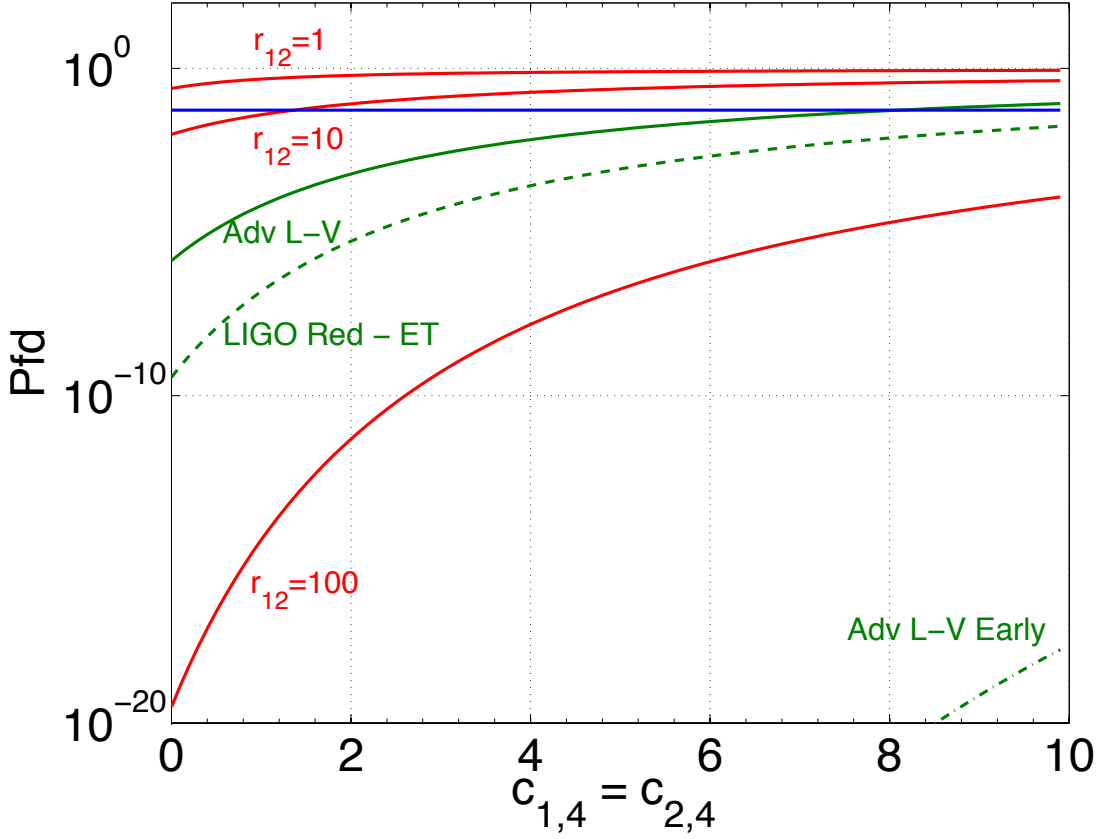}
\caption{Comparative efficiency of the CC and alternative detection
statistics. We take $T=10^{5}$ and $pfa=5\%$ and we assume that the signal
is Gaussian ($c_{3}=c_{4}=0$) and that the noise distributions are symmetric
($c_{1,3}=c_{2,3}=0$). Various red plots correspond to different values for
the ratio $r_{12}\equiv \dfrac{\protect\sigma _{1}^{2}}{\protect\sigma %
_{2}^{2}}=1,10,100,1000$. The parameter $\protect\alpha $ is chosen so that
the signal to noise ratio $SNR=\protect\sqrt{T}\dfrac{\protect\alpha ^{2}}{%
\protect\sigma _{2}\protect\sigma _{2}}=3.28$, a value that yields a $pfd=5\%
$ for the CC statistic in the homogeneous case $r_{12}=1$. When $%
c_{1,4}=c_{2,4}=0$, $pfd_{alt}\simeq 0.2,2\times 10^{-10},3\times 10^{-20}$
for $r_{12}=1,10,100$. The green plots correspond to realistic
values for the cross-correlation between Advanced LIGO and Advanced Virgo at
their nominal sensitivities ($r_{12}=48$) and between Einstein Telescope and
Advanced LIGO with the LIGO Red possible upgraded sensitivity ($r_{12}=30$).
We also considered an average value of $r_{12}=400$ corresponding to the
cross-correlation between Advanced LIGO and Advanced Virgo during the early
phases of the development of the detectors. When $c_{1,4}=c_{2,4}=0$, $%
pfd_{alt}\simeq 1\times 10^{-6},4\times 10^{-10},3\times 10^{-20},2\times
10^{-153}$ for $r_{12}=$ 30, 48 and 400.}
\label{fig1}
\end{figure}

\begin{figure}[tbp]
\includegraphics[width=\textwidth]{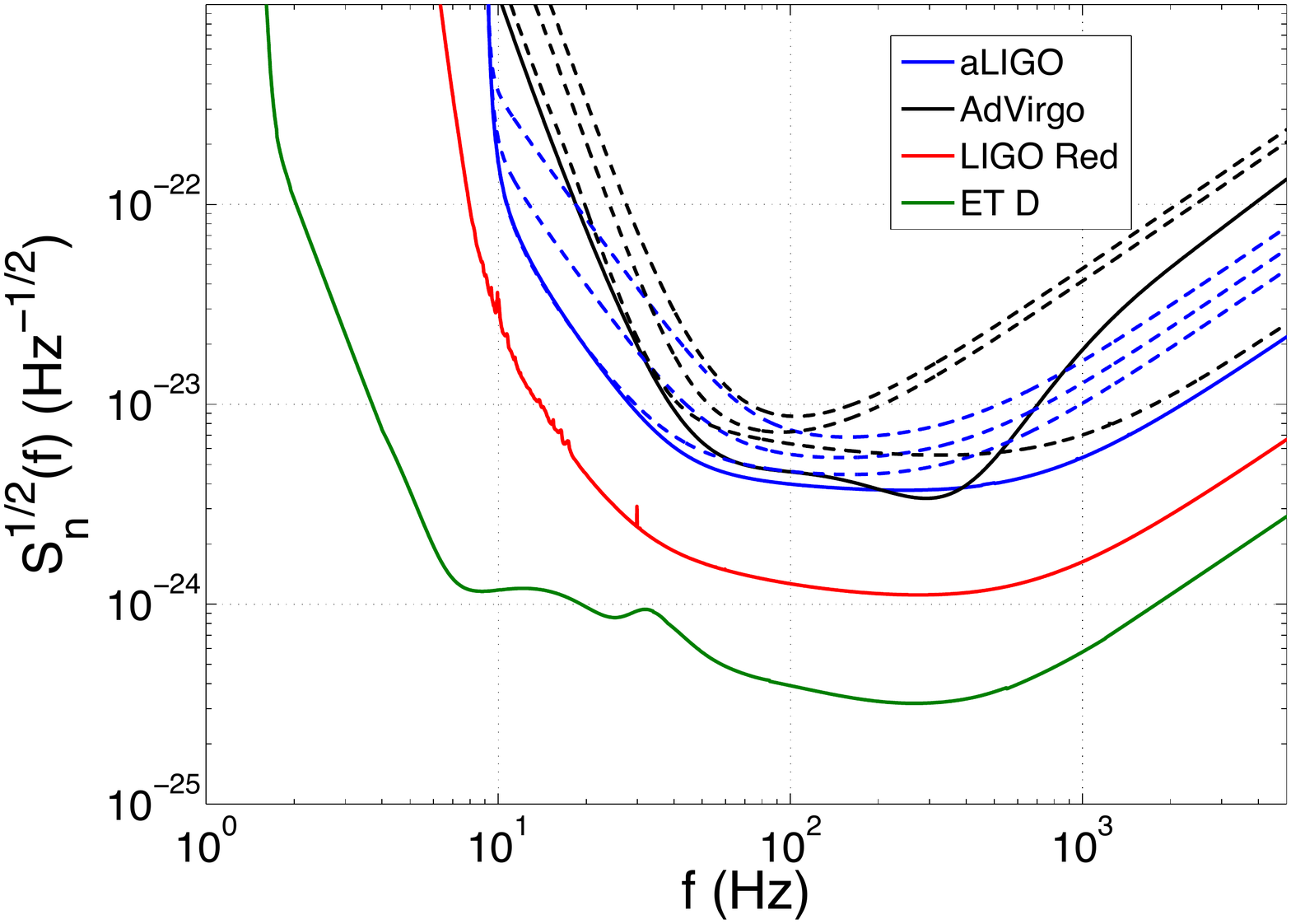}
\caption{Expected sensitivity of Advanced LIGO and Advanced Virgo (blue and
black continuous lines), LIGORed (dashed red line) and ET-D (green
continuous line). The evolution of the sensitivity during the Advanced LIGO
and Virgo early, middle and late phases are also shown in dashed blue and
black lines. }
\label{fig2}
\end{figure}

In Fig.~\ref{fig1}, we first plot the probability of a false dismissal ($pfd$%
) as a function of the 4th cumulant of the noise distribution, 
assumed to be identical for both detectors ($c_{1,4}=c_{2,4}$),
over a resonnable range of values expected for classical distributions such
as Gaussian, Hypersecant, Logistic or Laplace (see \cite{2014PhRvD..89l4009M}%
), for the CC statistic (in blue) and the alternative statistic (in red and
green). Here we assume that the signal is Gaussian ($c_{3}=c_{4}=0$) and
that the noise distributions are symmetric ($c_{1,3}=c_{2,3}=0$). 
The parameter $\alpha $ is chosen so that the signal-to-noise
ratio is $SNR=\sqrt{T}\dfrac{\alpha ^{2}}{\sigma _{1}\sigma _{2}}=3.28$,
 a value yielding a probability of false alarm $pfd=5\%$ 
 for the CC statistic in the homogeneous case where detectors 1 and 2 have
the same sensitivity, i.e. when the ratio between the detector noise
variances is $r_{12}\equiv \dfrac{\sigma _{1}^{2}}{\sigma _{2}^{2}}=1$. The various red plots indicate different values of this ratio, $%
r_{12}=1,10,100$, where detectors 1 and 2 are chosen so that $%
r_{12}>1$. The green plots correspond to realistic values for the
cross-correlation between Advanced LIGO and Advanced Virgo at their nominal
sensitivity ($r_{12}=48$), and between Einstein Telescope and LIGO
Red \cite{LIGORed}, a possible Advanced LIGO sensitivity upgrade ($r_{12}=30
$). We also considered a value $r_{12}=400$ corresponding
to the maximum expected cross-correlation between Advanced LIGO and Advanced
Virgo during the early phases of the development of the detectors \cite%
{phases}. The projected nominal and early sensitivities, in term of the
square root of the power spectral density $S_{n}$, of Advanced LIGO
and Advanced Virgo \cite{AdLIGO,AdVIRGO}, along with the LIGO Red noise
curve \cite{LIGORed} and the proposed Einstein Telescope sensitivity ET-D 
\cite{2011CQGra..28i4013H} are plotted on Fig.~\ref{fig2}. The corresponding
noise variances, calculated as $\sigma _{n}^{2}=\int_{f_{\min }}^{f_{\max
}}dfS_{n}(f)$, where $f_{\min }=10$ Hz and $f_{\max }= 250$ Hz is the typical frequency band used for the
cross-correlation analysis \cite{2015arXiv150606744M} are reported in Table~%
\ref{tab:Table1}.

\begin{table}[tbp]
\caption{Noise variance levels $\protect\sigma _{1}$ and $\protect\sigma _{2}
$, and ratio $r_{12}=\dfrac{\protect\sigma _{1}^{2}}{\protect\sigma _{2}^{2}}
$, calculated as $\protect\sigma _{n}^{2}=\protect\int_{f_{\min }}^{f_{\max
}}dfS_{n}(f)$, where $f_{\min }=10$ Hz and $f_{\max }=$ 250 Hz is the
typical frequency band used for the cross-correlation analysis, for Advanced
LIGO with Advanced Virgo (aLIGO and AdV) at design sensitivities and during
the early phases of development of the detectors (early, middle and late),
and for Einstein Telescope (ET-D sensitivity) with LIGO Red \protect\cite%
{LIGORed}, a possible Advanced LIGO sensitivity upgrade.}%
\begin{ruledtabular}
\begin{tabular}{llll}
Pair & $\sigma_1^2$ & $\sigma_2^2$  & $r_{12} $ \\
\hline
AdV -- aLIGO &$3.6 \times 10^{-44}$ &$1.7 \times 10^{-42}$  & \\
 \:\:  early-- middle (6 months) & $3.8 \times 10^{-41}$ & $1.0 \times 10^{-43}$  &  371 \\
 \:\:  middle -- late (9 months) & $1.5 \times 10^{-41}$ &  $3.6 \times 10^{-44}$  & 402 \\
 \:\:  late -- design (12 months) &$3.2 \times 10^{-42}$ & $3.6 \times 10^{-44}$ &  88 \\
\hline
LIGO Red -- ET-D& $6.0 \times 10^{-47}$ & $1.9 \times 10^{-45}$ & 31\\
\end{tabular}
\end{ruledtabular}
\label{tab:Table1}
\end{table}

We confirm that $pfd_{CC}=5\%$ when the two detectors have equal sensitivity
($r_{12}=1$) and we find that $pfd_{CC}<pfd_{alt}$ in this case, as
expected. On the other hand, as we let the ratio $r_{12}$ increase, we find
that the probability of a false dismissal for the CC statistic decreases
very slightly, while the probability of a false dismissal for the
alternative statistic decreases very fast. As a result, we have that $%
pfd_{CC}>pfd_{alt}$ when $r_{12}>>10$. Note that $pfd_{alt}$ is almost $0$
when $r_{12}=100$, a situation in which the signal appears large compared to
the noise in the most sensitive detector. We therefore obtain that the
standard CC statistic can be dominated by the alternative statistic when
detector sensitivities exhibit substantial differences, even when both
signal and noise distributions are Gaussian. In fact, the domination of the
alternative statistic decreases as $c_{1,4}=c_{2,4}$ increases. This can be
explained by the fact that increases in $c_{1,4}$ and $c_{2,4}$ lead to
increase in the variance of the alternative detection statistic, which is
detrimental to the performance of the detection methodology. We
notice that for the values of $r_{12}$ expected for the current
and next generations of detectors, the alternative statistic usually
performs better than the CC statistic, except for the most pessimistic case
when $c_{1,4}=c_{2,4}\geq 8$ for the cross-correlation between
LIGO Red and ET ($r_{12}\sim 48$).

\begin{figure}[tbp]
\includegraphics[width=\textwidth]{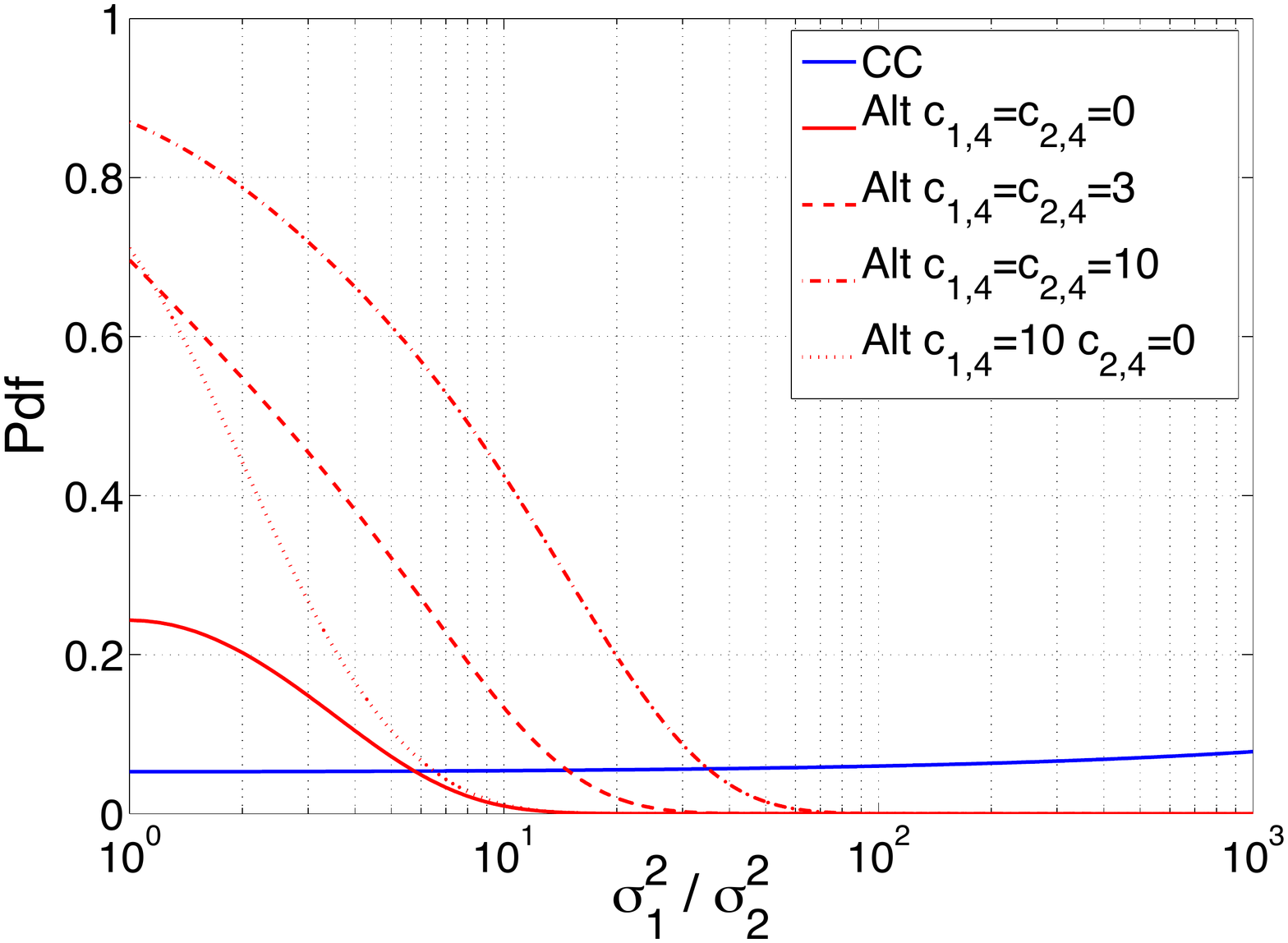}
\caption{Probability of a false alarm for the CC and alternative detection
statistics as a function of the ratio of detector sensitivities. We take $%
T=10^{5}$ and $pfa=5\%$ and we assume that the signal is Gaussian ($%
c_{3}=c_{4}=0$) and that the noise distributions are symmetric ($%
c_{1,3}=c_{2,3}=0$). Various red plots correspond to different choices for
the pair of parameters $c_{1,4}$ and $c_{2,4}$. The parameter $\protect%
\alpha $ is chosen so that the signal to noise ratio $SNR=\protect\sqrt{T}%
\dfrac{\protect\alpha ^{2}}{\protect\sigma _{2}\protect\sigma _{2}}=3.28$, a
value that yields a $pfd=5\%$ for the CC statistic in the homogeneous case $%
r_{12}\equiv \dfrac{\protect\sigma _{1}^{2}}{\protect\sigma _{2}^{2}}=1$.}
\label{fig3}
\end{figure}

In Fig.~\ref{fig3}, we consider the dual perspective, where the probability
of a false dismissal is plotted against the ratio $r_{12}$ for typical
values of $c_{1,4}=c_{2,4}=0$, 3 and 10. Again, as the ratio increases, we
find a small deterioration in the performance of the CC statistic (the probability of false dismissal increases from $pfd_{cc}\simeq 5\%$ for $r_{12}=1$ to $pfd_{cc}\simeq 8\%$ for $%
r_{12}=1000$) and a substantial improvement for the alternative statistic.
In the case of a normally distributed noise process ($%
c_{1,4}=c_{2,4}=0$), the alternative statistic outperforms the CC
statistic for values $r_{12}\geq 6$ and the probability of a false
dismissal becomes negligible, as small as $pfd_{alt}\sim 4\times 10^{-10}$ for the pair Advanced LIGO and Advanced Virgo ($r_{12}\simeq 48$) and $pfd_{alt}\sim 1\times 10^{-6}$ for the pair LIGO
Red and ET ($r_{12}\simeq 30$), which translates into a gain of 8
and 4 orders of magnitude, respectively, compared to the CC statistic. For
the cross-correlation between Advanced LIGO and Advanced Virgo during the
early stages of development ($r_{12}\simeq 400$), the probability
of a false dismissal is almost zero ($pfd_{alt}\sim 2\times 10^{-153}$).

So as to better understand why the presence of heterogenous detectors has
such a strong impact on the relative efficiency of the CC versus alternative
detection statistic, we consider two contrasted situations, a homogenous
case situation (C1), with $\sigma _{1}=\sigma _{2}=\sigma $ and a
heterogenous case situation (C2), with $\sigma _{1}=\dfrac{1}{\sigma }$ and $%
\sigma _{2}=\sigma ^{3}$. We note that by construction the product $\sigma
_{1}^{2}\sigma _{2}^{2}=\sigma ^{4}$ in both cases but the sum $\sigma
_{1}^{2}+\sigma _{2}^{2}$ is different and substantially higher in the
heterogenous case situation.

Assuming Gaussian distributions for signal and noise, we have:%
\begin{eqnarray*}
\mu _{CC}^{ns} &=&0\text{, yielding the same value in C1 and C2} \\
\mu _{alt}^{ns} &=&\sigma _{1}^{2}\sigma _{2}^{2}\text{, yielding the same
value in C1 and C2} \\
\sigma _{CC}^{2,ns} &=&\dfrac{\sigma _{1}^{2}\sigma _{2}^{2}}{T}\text{,
yielding the same value in C1 and C2} \\
\sigma _{alt}^{2,ns} &=&\dfrac{8\sigma _{1}^{4}\sigma _{2}^{4}}{T}\text{,
yielding the same value in C1 and C2}
\end{eqnarray*}%
and we also have:%
\begin{eqnarray*}
\mu _{CC} &=&\alpha ^{2}\text{, yielding the same value in C1 and C2} \\
\mu _{alt} &=&\sigma _{1}^{2}\sigma _{2}^{2}+\alpha ^{2}\left( \sigma
_{1}^{2}+\sigma _{2}^{2}\right) +3\alpha ^{4}+c_{4}\text{, yielding a
greater value for C2} \\
\sigma _{CC}^{2} &=&\dfrac{1}{T}\left( \sigma _{1}^{2}\sigma _{2}^{2}+\alpha
^{2}\left( \sigma _{1}^{2}+\sigma _{2}^{2}\right) +2\alpha ^{4}+c_{4}\right) 
\text{, yielding a greater value for C2} \\
\sigma _{alt}^{2} &=&\text{ expression given in Eq.~\ref{varalt}, yielding a
greater value for C2}
\end{eqnarray*}

As a result, we find that the probability of a false alarm is the same in C1
and C2 for both the standard CC and the alternative detection statistics,
since this probability only depends on the distributions of the detection
statistics in the absence of a signal, which are the same for C1 and C2. In
the presence of a signal, we find for the alternative statistic that moving
from the homogenous case (C1) to the heterogeneous case (C2) leads to an
increase in the mean, which has a positive impact on sensitivity, and an
increase in the variance, which has a negative impact on sensitivity.
Overall, the net effect is positive, as can be seen from Fig.~\ref{fig2} and %
\ref{fig3}. For the cross-correlation statistic, the mean is not impacted
but there is an increase in variance, which is detrimental to detection.
Overall, we confirm that the case with heterogeneous sensitivities is more
favorable for the alternative statistic than it is for the CC detection
statistic.

\begin{figure}[tbp]
\includegraphics[width=\textwidth]{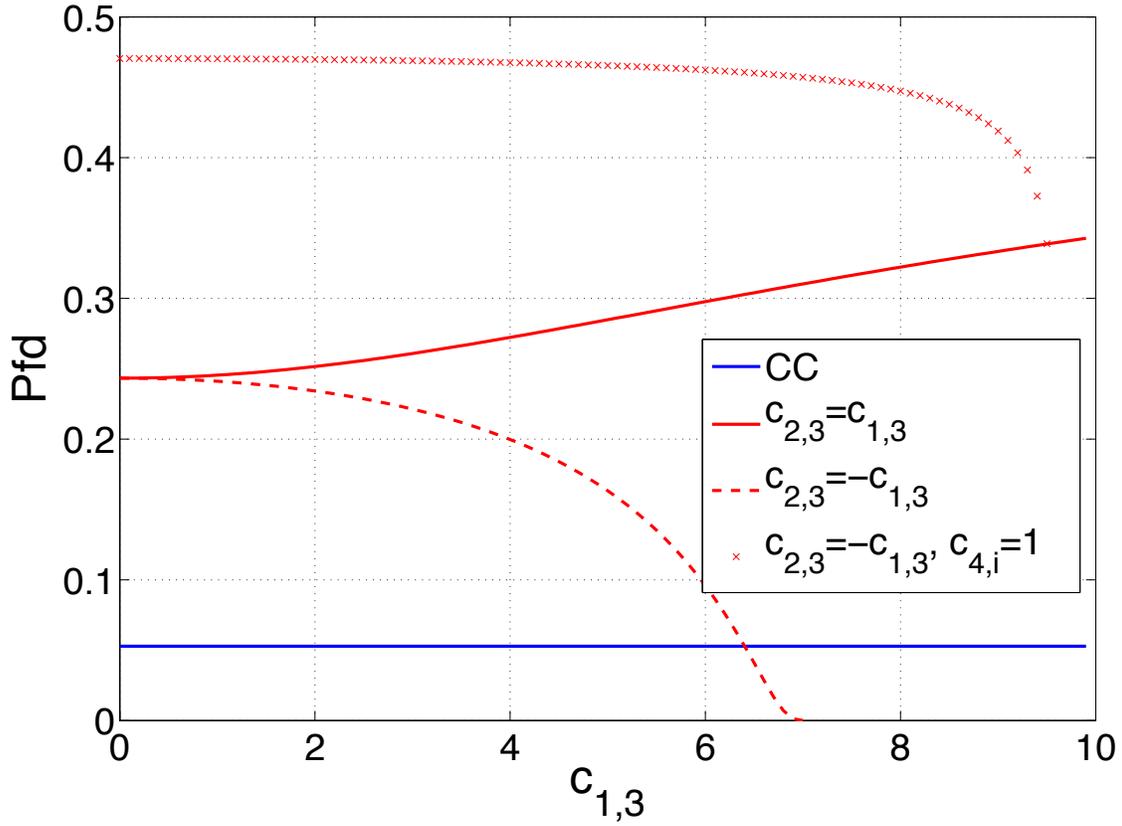}
\caption{Impact of the skewness of the detector noise on the relative
efficiency of the CC and alternative detection statistics for homogeneous
detector sensitivities.. We take $T=10^{5}$ and $pfa=5\%$ and we assume that
the signal is Gaussian ($c_{3}=c_{4}=0$). We consider the homogeneous case $%
r_{12}\equiv \dfrac{\protect\sigma _{1}^{2}}{\protect\sigma _{2}^{2}}=1 $.
The parameter $\protect\alpha $ is chosen so that the signal to noise ratio $%
SNR=\protect\sqrt{T}\dfrac{\protect\alpha ^{2}}{\protect\sigma _{2}\protect%
\sigma _{2}}=3.28$, a value that yields a $pfd=5\%$ for the CC statistic for 
$\dfrac{\protect\sigma _{1}^{2}}{\protect\sigma _{2}^{2}}=1$. Various red
plots correspond to different choices of value for $c_{1,3}$ and $c_{2,3}$%
,~with $c_{4,i}=0$ for $i=1,2$, except for the crossed red line, where it is
equal to 1.}
\label{fig4}
\end{figure}

We now turn to the analysis of the impact of the 3rd moment of the noise
distribution, which has been assumed to be zero so far. In Fig.~\ref{fig4},
we show the probability of a false dismissal for the CC statistic (in blue)
and the alternative statistic (in red) as a function of the third-order
cumulant of the distribution of the noise for the first detector. We take $%
T=10^{5}$ and $pfa=5\%$ and we assume that the signal is Gaussian ($%
c_{3}=c_{4}=0$). We consider the homogeneous case $r_{12}\equiv \dfrac{%
\sigma _{1}^{2}}{\sigma _{2}^{2}}=1$. As usual, the parameter $\alpha $ is
chosen so that the signal-to-noise ratio $SNR=\sqrt{T}\dfrac{\alpha ^{2}}{%
\sigma _{2}\sigma _{2}}=3.28$, a value that yields a $pfd=5\%$ for the CC
statistic for $r_{12}=1$. Various red plots correspond to different values
for $c_{2,3}$,~with $c_{4,i}=0$ for $i=1,2$, except for the crossed red
line, where it is equal to 1. We find that the alternative statistic may
dominate the CC statistic when $c_{1,3}$ and $c_{2,3}$ are of opposite
signs. Indeed, the 3rd higher-order cumulants of the noise distributions do
not impact the mean and variance of the CC statistic, but have an impact on
the variance of the alternative statistic. Unlike the 4th order cumulant
that is always positive, the 3rd higher-order cumulants can principle take
on any negative or positive value, and taking them of opposite signs leads
to a reduction in the variance of the alternative detection statistic.
Indeed, these parameters enter the expression for the variance of the
alternative statistic in Eq.~\ref{varaltsym} through the term $+16\mu
_{2}\mu _{1,3}\mu _{2,3}$, which is negative when noise distributions have
skewness parameters of opposite signs, suggesting a diversification effect
(note that $\mu _{3}=c_{3}$ for centered distributions). On the other hand,
we note again that an increase in $c_{4,i} $ (crossed red line in Fig.~\ref%
{fig3}) leads to a deterioration of the performance of the alternative
statistic.

\begin{figure}[tbp]
\includegraphics[width=\textwidth]{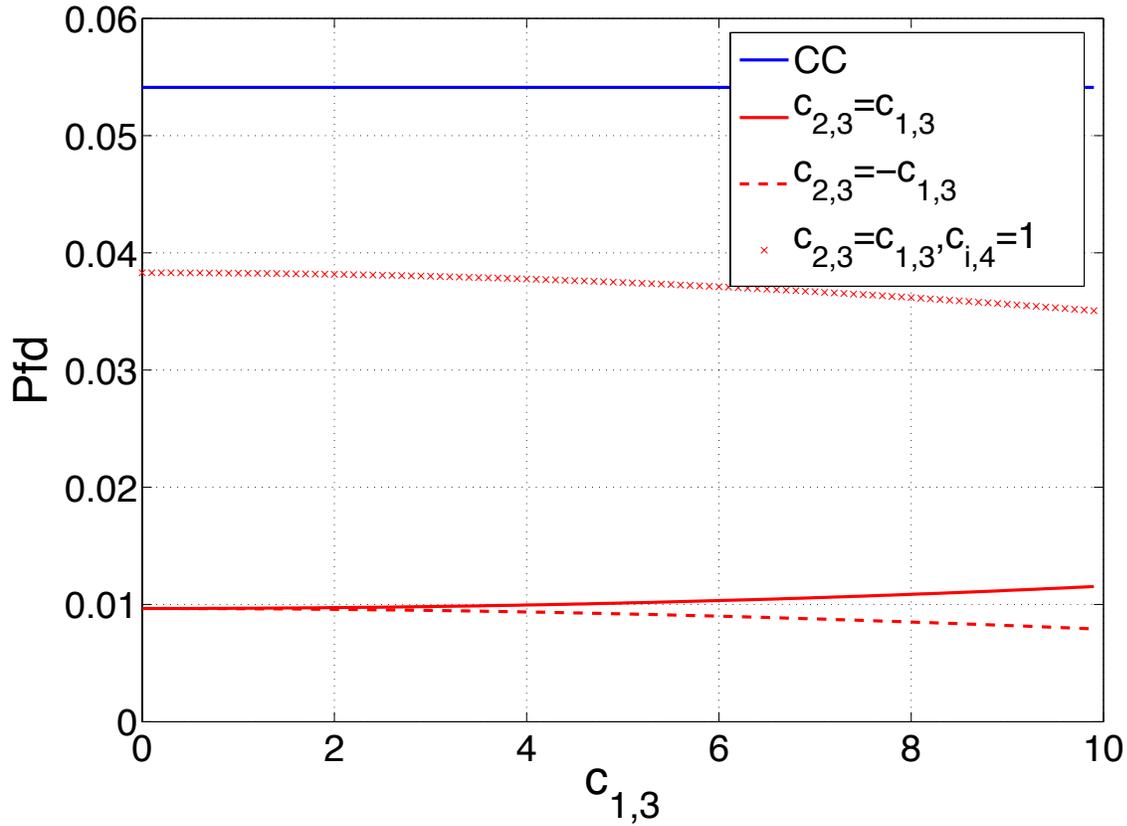}
\caption{Impact of the skewness of the detector noise on the relative
efficiency of the CC and alternative detection statistics for heterogeneous
detector sensitivities. We take $T=10^{5}$ and $pfa=5\%$ and we assume that
the signal is Gaussian ($c_{3}=c_{4}=0$). We consider the homogeneous case $%
r_{12}\equiv \dfrac{\protect\sigma _{1}^{2}}{\protect\sigma _{2}^{2}}=10$.
The parameter $\protect\alpha $ is chosen so that the signal to noise ratio $%
SNR=\protect\sqrt{T}\dfrac{\protect\alpha ^{2}}{\protect\sigma _{2}\protect%
\sigma _{2}}=3.28$, a value that yields a $pfd=5\%$ for the CC statistic for 
$r_{12}=1$. Various red plots correspond to different choices of value for $%
c_{1,3}$ and $c_{2,3}$,~with $c_{4,i}=0$ for $i=1,2$, except for the crossed
red line, where it is equal to 1. }
\label{fig5}
\end{figure}

In Fig.~\ref{fig5}, we repeat the analysis but focus on the case of
heterogenous detector sensitivities by taking $r_{12}\equiv \dfrac{\sigma
_{1}^{2}}{\sigma _{2}^{2}}=10$, while we had $r_{12} =1$ in Fig.~\ref{fig4}.

We find again that the alternative statistic dominates the CC statistic, an
effect that is stronger when $c_{1,3}$ and $c_{2,3}$ are of opposite signs,
but decreases when $c_{4,i}$ increases.

\section{Estimation Methods for Non-Gaussian Signal and Non-Gaussian Noise
Distributions}

The analysis in the previous Section suggests that the use of the
alternative statistic may lead to noticeable sensitivity gains when noise
distributions are non-Gaussian and/or when detector sensitivities exhibit
substantial differences. It should be noted, however, that using the
alternative statistic requires in the non-Gaussian case the use of robust
estimates not only for the variance, but also the skewness and kurtosis, of
the signal and noise distributions. In what follows, we show how to obtain
such estimates by extending standard maximum likelihood estimation
methodologies to situations involving possibly non-Gaussian signal and noise
distributions. As such, these results generalize early results reported by 
\cite{2014PhRvD..89l4009M} who have focussed on a situation involving a
non-Gaussian signal distribution, while maintaining the assumption of a
Gaussian noise distribution.

We denote by $f_{n_{i}}$ and $f_{s}$, respectively, the density function for
the noise and the signal:%
\begin{eqnarray*}
\Pr \left( \mathcal{N}_{i}\in \left[ n,n+dn\right] \right)
&=&f_{n_{i}}\left( n\right) dn \\
\Pr \left( \mathcal{S}\in \left[ s,s+sn\right] \right) &=&f_{s}\left(
s\right) ds
\end{eqnarray*}

We also denote by $f_{n}\equiv f_{n}\left( n_{1t},n_{2t}\right) _{t=1,...,T}$
the joint probability distribution for the noise in the two detectors. The
standard Bayesian approach for signal detection consists in finding the
value for the unknown parameters so as to minimize the false dismissal
probability at a fixed value of the false alarm probability. This criteria,
known as the Neyman-Pearson criteria, is uniquely defined in terms of the
so-called likelihood ratio $\Lambda $ given by:%
\begin{equation*}
\Lambda =\frac{\left. p_{h}\right\vert _{X=1}}{\left. p_{h}\right\vert _{X=0}%
} 
\end{equation*}%
where $\left. p_{h}\right\vert _{X=1}$ (respectively, $\left.
p_{h}\right\vert _{X=0}$) is the conditional density for the measurement
output if a signal is present (respectively, absent). A natural
approximation of the likelihood ratio is the maximum likelihood detection
statistic defined by \cite{2003PhRvD..67h2003D}: 
\begin{equation}
\Lambda _{ML}=\frac{\underset{\alpha ,\sigma _{1},\sigma _{2}}{\max }\int
\left. f_{s}\right\vert _{X=1}\left( s\right) \left. f_{n}\right\vert
_{X=1}\left( h-s\right) ds}{\underset{\sigma _{1},\sigma _{2}}{\max }\left.
f_{n}\right\vert _{X=0}\left( h\right) }  \label{ML}
\end{equation}%
and the maximum likelihood estimators for the unknown signal and noise
standard-deviation parameters $\alpha ,$ $\sigma _{1}$ and $\sigma _{2}$ are
given as the corresponding likelihood maximizing quantities.

\subsection{Full Gaussian Case}

It is typically assumed that both the noise and signal are normally
distributed, $f_{n}$ and $f_{s}$ are Gaussian probability distribution
functions, that is we may assume:%
\begin{eqnarray*}
f_{n_{i}}\left( n_{it}\right) &=&\frac{1}{\sqrt{2\pi }\sigma _{i}}e^{-\frac{%
\left( n_{it}-\mu _{i}\right) ^{2}}{2\sigma _{i}^{2}}}\text{ for }i=1,2 \\
f_{s}\left( s_{t}\right) &=&\frac{1}{\sqrt{2\pi }\alpha }e^{-\frac{\left(
s_{t}-\beta \right) ^{2}}{2\alpha ^{2}}}
\end{eqnarray*}%
where we also assume that both the noise and signal are weakly stationary
processes so that their moments are constant through time. We further assume
the noise in detector one and two are uncorrelated with a zero mean for both
detectors. Under these assumptions, we have:%
\begin{equation*}
f_{n}\left( n_{1t},n_{2t}\right) =\frac{1}{2\pi \sigma _{1}\sigma _{2}}e^{-%
\frac{n_{1t}^{2}}{2\sigma _{1}^{2}}-\frac{n_{2t}^{2}}{2\sigma _{2}^{2}}} 
\end{equation*}%
and finally, assuming zero serial correlation: 
\begin{eqnarray*}
f_{n} &\equiv &f_{n}\left( n_{1t},n_{2t}\right)
_{t=1,...,T}=\prod\limits_{t=1}^{T}\frac{1}{2\pi \sigma _{1}\sigma _{2}}e^{-%
\frac{n_{1t}^{2}}{2\sigma _{1}^{2}}-\frac{n_{2t}^{2}}{2\sigma _{2}^{2}}} \\
f_{s} &\equiv &f_{s}\left( s_{t}\right) _{t=1,...,T}=\prod\limits_{t=1}^{T}%
\frac{1}{\sqrt{2\pi }\alpha }e^{-\frac{\left( s_{t}-\beta \right) ^{2}}{%
2\alpha ^{2}}}
\end{eqnarray*}

Typically, one also assume that the mean value $\beta $ for the signal is
zero so that the unknown parameters are $\alpha ,$ $\sigma _{1}$ and $\sigma
_{2}$. Then, we have that the denominator of equation (\ref{ML}) is given by:%
\begin{equation*}
\underset{\sigma _{1},\sigma _{2}}{\max }\left. f_{n}\right\vert
_{X=0}\left( h\right) =\underset{\sigma _{1},\sigma _{2}}{\max }%
\prod\limits_{t=1}^{T}\frac{1}{2\pi \sigma _{1}\sigma _{2}}e^{-\frac{%
h_{1t}^{2}}{2\sigma _{1}^{2}}-\frac{h_{2t}^{2}}{2\sigma _{2}^{2}}}=\underset{%
\sigma _{1},\sigma _{2}}{\max }\frac{1}{\left( 2\pi \sigma _{1}\sigma
_{2}\right) ^{T}}\exp \left[ -\sum\limits_{t=1}^{T}\frac{h_{1t}^{2}}{2\sigma
_{1}^{2}}-\sum\limits_{t=1}^{T}\frac{h_{2t}^{2}}{2\sigma _{2}^{2}}\right] 
\end{equation*}

Introducing for $i=1,2$: 
\begin{equation*}
\overline{\sigma }_{i}^{2}=\frac{1}{T}\sum\limits_{t=1}^{T}h_{it}^{2} 
\end{equation*}%
we finally have that:%
\begin{equation}
\underset{\sigma _{1},\sigma _{2}}{\max }f_{n}=\underset{\sigma _{1},\sigma
_{2}}{\max }\frac{1}{\left( 2\pi \sigma _{1}\sigma _{2}\right) ^{T}}\exp %
\left[ -\frac{T}{2}\left( \frac{\overline{\sigma }_{1}^{2}}{\sigma _{1}^{2}}+%
\frac{\overline{\sigma }_{2}^{2}}{\sigma _{2}^{2}}\right) \right]
\label{den}
\end{equation}

It is straightforward to see that the maximum for equation (\ref{den}) is
reached for $\sigma _{i}^{2}=\overline{\sigma }_{i}^{2}$ and that maximum is
given by:%
\begin{equation*}
\underset{\sigma _{1},\sigma _{2}}{\max }f_{n}=\frac{1}{\left( 2\pi 
\overline{\sigma }_{1}\overline{\sigma }_{2}\right) ^{T}}\exp \left[ -\frac{T%
}{2}\left( 1+1\right) \right] =\frac{1}{\left( 2\pi \overline{\sigma }_{1}%
\overline{\sigma }_{2}\right) ^{T}}\exp \left( -T\right) 
\end{equation*}

Finally, we obtain the following expression:%
\begin{eqnarray*}
\Lambda _{ML} &=&\frac{\underset{\alpha ,\sigma _{1},\sigma _{2}}{\max }\int
f_{s}\left( s\right) f_{n}\left( h-s\right) ds}{\underset{\sigma _{1},\sigma
_{2}}{\max }f_{n}} \\
&=&\left( 2\pi \overline{\sigma }_{1}\overline{\sigma }_{2}\right) ^{T}\exp
\left( T\right) \underset{\alpha ,\sigma _{1},\sigma _{2}}{\max }%
\prod\limits_{t=1}^{T}\int_{-\infty }^{+\infty }f_{s}\left( s_{t}\right) 
\frac{1}{2\pi \sigma _{1}\sigma _{2}}\exp \left[ -\frac{\left(
h_{1t}-s_{t}\right) ^{2}}{2\sigma _{1}^{2}}-\frac{\left( h_{2t}-s_{t}\right)
^{2}}{2\sigma _{2}^{2}}\right] ds_{t} \\
&=&\underset{\alpha ,\sigma _{1},\sigma _{2}}{\max }\prod\limits_{t=1}^{T}%
\frac{\overline{\sigma }_{1}\overline{\sigma }_{2}}{\sigma _{1}\sigma _{2}}%
\int_{-\infty }^{+\infty }f_{s}\left( s_{t}\right) \exp \left[ -\frac{\left(
h_{1t}-s_{t}\right) ^{2}}{2\sigma _{1}^{2}}-\frac{\left( h_{2t}-s_{t}\right)
^{2}}{2\sigma _{2}^{2}}+1\right] ds_{t}
\end{eqnarray*}

We now specialize the analysis of the specific situation where the signal
has a Gaussian distribution. In this case, and maintaining the assumption
that the mean value for the signal is zero, we obtain:%
\begin{equation*}
f_{s}\equiv f_{s}\left( s_{t}\right) _{t=1,...,T}=\prod\limits_{t=1}^{T}%
\frac{1}{\sqrt{2\pi }\alpha }e^{-\frac{s_{t}^{2}}{2\alpha ^{2}}} 
\end{equation*}

We thus have:%
\begin{equation*}
\Lambda _{ML}=\underset{\alpha ,\sigma _{1},\sigma _{2}}{\max }%
\prod\limits_{t=1}^{T}\frac{\overline{\sigma }_{1}\overline{\sigma }_{2}}{%
\sqrt{2\pi }\alpha \sigma _{1}\sigma _{2}}\int_{-\infty }^{+\infty }\exp %
\left[ -\frac{s_{t}^{2}}{2\alpha ^{2}}-\frac{\left( h_{1t}-s_{t}\right) ^{2}%
}{2\sigma _{1}^{2}}-\frac{\left( h_{2t}-s_{t}\right) ^{2}}{2\sigma _{2}^{2}}%
+1\right] ds_{t} 
\end{equation*}

After tedious calculations, one obtains (see \cite{2003PhRvD..67h2003D}): 
\begin{equation}
\Lambda _{ML}=\underset{\alpha ,\sigma _{1},\sigma _{2}\geq 0}{\max }\left\{ 
\frac{\overline{\sigma }_{1}\overline{\sigma }_{2}}{\sqrt{\sigma
_{1}^{2}\sigma _{2}^{2}+\sigma _{1}^{2}\alpha ^{2}+\sigma _{2}^{2}\alpha ^{2}%
}}\exp \left[ \frac{\frac{\overline{\sigma }_{1}^{2}}{\sigma _{1}^{4}}+\frac{%
\overline{\sigma }_{2}^{2}}{\sigma _{2}^{4}}+\frac{2\overline{\alpha }^{2}}{%
\sigma _{1}^{2}\sigma _{2}^{2}}}{2\left( \frac{1}{\sigma _{1}^{2}}+\frac{1}{%
\sigma _{2}^{2}}+\frac{1}{\alpha ^{2}}\right) }-\frac{\overline{\sigma }%
_{1}^{2}}{2\sigma _{1}^{2}}-\frac{\overline{\sigma }_{2}^{2}}{2\sigma
_{2}^{2}}+1\right] \right\} ^{T}
\end{equation}

One can show that the maximum is reached for:%
\begin{eqnarray*}
\alpha ^{2} &=&\widehat{\alpha }^{2}\equiv \left( \overline{\alpha }%
^{2}\right) ^{+} \\
\sigma _{i}^{2} &=&\widehat{\sigma }_{i}^{2}\equiv \left( \overline{\sigma }%
_{i}^{2}-\widehat{\alpha }^{2}\right) ^{+}
\end{eqnarray*}%
where $\left( x\right) ^{+}=x$ if $x>0$ and $\left( x\right) ^{+}=0$
otherwise, which arise because of the positivity constraints on $\alpha ,$ $%
\sigma _{1}$ and $\sigma _{2}$ in he maximization procedure.

The corresponding detection statistic is:%
\begin{equation}
\Lambda _{ML}^{G}=\left( 1-\frac{\widehat{\alpha }^{4}}{\overline{\sigma }%
_{1}^{2}\overline{\sigma }_{2}^{2}}\right) ^{-T/2}
\end{equation}

The cross-correlation statistic $\Lambda _{cc}^{G}$ can be obtained from $%
\Lambda _{ML}^{G}$ via a monotonic transformation which preserves false
dismissal versus false alarm curves (see again \cite{2003PhRvD..67h2003D}): 
\begin{equation}
\Lambda _{cc}^{G}=\sqrt{1-\left( \Lambda _{ML}^{G}\right) ^{-2/T}}=\frac{%
\widehat{\alpha }^{2}}{\overline{\sigma }_{1}\overline{\sigma }_{2}}
\end{equation}

\subsection{Gaussian Signal and Non-Gaussian Noise}

In \cite{2014PhRvD..89l4009M}, the Gaussian assumption was maintained for
the detector noise distribution, but relaxed for the signal distribution. In
what follows, we consider the opposite situation, namely a normally
distributed signal, and a potentially non-Gaussian noise distribution. In
other words, we assume:

\begin{eqnarray*}
f_{s}\left( s_{t}\right) &=&\frac{1}{\sqrt{2\pi }\alpha }e^{-\frac{s_{t}^{2}%
}{2\alpha ^{2}}}\text{ } \\
f_{n_{i}}\left( n_{it}\right) &\neq &\frac{1}{\sqrt{2\pi }\sigma _{i}}e^{-%
\frac{n_{it}^{2}}{2\sigma _{i}^{2}}}\text{ for }i=1,2
\end{eqnarray*}

As in \cite{2014PhRvD..89l4009M}, we propose to use a semi-parametric
approach which allows one to approximate the unknown density as a
transformation of a reference function (typically the Gaussian density),
involving higher-order moments/cumulants of the unknown distribution. This
approach has been heavily used in statistical problems involving a mild
departure from the Gaussian distribution. In what follows, we will show that
it allows us to obtain an analytical derivation of the nearly optimal
maximum likelihood detection statistics for non-Gaussian gravitational wave
stochastic backgrounds.

We want to approximate $f_{n_{i}}$, the density function of the unknown
distribution of the noise distribution $\mathcal{N}_{i}$, as a function of
the Gaussian density function $f_{n_{i}}^{G}\left( x\right) $ and a
multiplicative deviation $g_{n_{i}}\left( x\right) $ from the Gaussian
density function. To achieve this objective, we use the \textit{Edgeworth
expansion}, which is based on the assumption that the unknown signal
distribution is the sum of normalized i.i.d. (non necessarily Gaussian)
variables. In other words, it provides asymptotic correction terms to the
Central Limit Theorem up to an order that depends on the number of moments
available. When taken to the fourth-order level, the Edgeworth expansion
reads as follows (see for example \cite{feller2008introduction} (1971, P.
535) for the proof, and additional results regarding the convergence rate of
the Edgeworth expansion): 
\begin{equation}
f_{n_{i}}\left( x\right) \simeq f_{n_{i}}^{G}\left( x\right) \left[ 1+\frac{%
c_{i,3}}{6\sigma _{i}^{3}}H_{3}\left( \frac{x}{\sigma _{i}}\right) +\frac{%
c_{i,4}}{24\sigma _{i}^{4}}H_{4}\left( \frac{x}{\sigma _{i}}\right) +\frac{%
c_{i,3}^{2}}{72\sigma _{i}^{6}}H_{6}\left( \frac{x}{\sigma _{i}}\right) %
\right]
\end{equation}%
where the 6th Hermite polynomial is defined as $H_{6}\left( x\right)
=x^{6}-15x^{4}+45x^{2}-15$. We finally have $f_{n_{i}}\left( x\right) \simeq
f_{n_{i}}^{G}\left( x\right) g_{n_{i}}\left( x\right) $ with:%
\begin{eqnarray}
f_{n_{i}}^{G}\left( x\right) &\equiv &\frac{1}{\sqrt{2\pi }\sigma _{i}}\exp %
\left[ -\frac{x^{2}}{2\sigma _{i}^{2}}\right] \\
g_{n_{i}}\left( x\right) &\equiv &\underset{b_{i,0}}{\underbrace{\left( 1+%
\frac{c_{i,4}}{8\sigma _{i}^{4}}-\frac{5c_{i,3}^{2}}{24\sigma _{i}^{6}}%
\right) }}\underset{b_{i,1}}{\underbrace{-\frac{c_{i,3}}{2\sigma _{i}^{4}}}}x%
\underset{b_{i,2}}{+\underbrace{\left( \frac{15c_{i,3}^{2}}{24\sigma _{i}^{8}%
}-\frac{c_{i,4}}{4\sigma _{i}^{6}}\right) }}x^{2}+\underset{b_{i,3}}{%
\underbrace{\frac{c_{i,3}}{6\sigma _{i}^{6}}}}x^{3}+\underset{b_{i,4}}{%
\underbrace{\left( \frac{c_{i,4}}{24\sigma _{i}^{8}}-\frac{5c_{i,3}^{2}}{%
24\sigma _{i}^{10}}\right) }}x^{4}+\underset{b_{i,6}}{\underbrace{\frac{%
c_{i,3}^{2}}{72\sigma _{i}^{6}}}}x^{6}
\end{eqnarray}

In this context, the likelihood maximization problem becomes:

\begin{eqnarray*}
\Lambda _{ML} &=&\underset{\alpha ,\sigma _{1},\sigma
_{2},c_{1,3},c_{1,4},c_{2,3},c_{2,4}}{\max }\prod\limits_{t=1}^{T}\frac{%
\overline{\sigma }_{1}\overline{\sigma }_{2}}{\sqrt{2\pi }\alpha \sigma
_{1}\sigma _{2}}\int_{-\infty }^{+\infty }\exp \left[ -\frac{s_{t}^{2}}{%
2\alpha ^{2}}-\frac{\left( h_{1t}-s_{t}\right) ^{2}}{2\sigma _{1}^{2}}-\frac{%
\left( h_{2t}-s_{t}\right) ^{2}}{2\sigma _{2}^{2}}+1\right] g_{n_{1}}\left(
x\right) g_{n_{2}}\left( x\right) ds_{t} \\
&=&\underset{\alpha ,\sigma _{1},\sigma _{2},c_{1,3},c_{1,4},c_{2,3},c_{2,4}}%
{\max }\prod\limits_{t=1}^{T}\frac{\sigma }{\alpha }\frac{\overline{\sigma }%
_{1}\overline{\sigma }_{2}}{\sigma _{1}\sigma _{2}}\exp \left[ -\frac{%
h_{1t}^{2}}{2\sigma _{1}^{2}}-\frac{h_{2t}^{2}}{2\sigma _{2}^{2}}+1\right]
\exp \left[ \frac{1}{2}\sigma ^{2}\left( \frac{h_{1t}}{\sigma _{1}^{2}}+%
\frac{h_{2t}}{\sigma _{2}^{2}}\right) ^{2}\right] \\
&&\times \int_{-\infty }^{+\infty }\frac{1}{\sigma \sqrt{2\pi }}\exp \left[ -%
\frac{1}{2\sigma ^{2}}\left( s_{t}-\mu _{t}\right) ^{2}\right]
g_{n_{1}}\left( x\right) g_{n_{2}}\left( x\right) ds_{t}
\end{eqnarray*}%
with%
\begin{eqnarray*}
\sigma &=&\left( \frac{1}{\alpha ^{2}}+\frac{1}{\sigma _{1}^{2}}+\frac{1}{%
\sigma _{2}^{2}}\right) ^{-\frac{1}{2}} \\
\mu _{t} &=&\left( \frac{h_{1t}}{\sigma _{1}^{2}}+\frac{h_{2t}}{\sigma
_{2}^{2}}\right) \sigma ^{2}
\end{eqnarray*}

Focussing for simplicity of exposure on symmetric noise distribution
functions (therefore such that $c_{i,3}=0$), we have:%
\begin{eqnarray*}
g_{n_{1}}\left( x\right) g_{n_{2}}\left( x\right) &=&\left(
b_{1,0}+b_{1,2}x^{2}+b_{1,4}x^{4}\right) \left(
b_{2,0}+b_{2,2}x^{2}+b_{2,4}x^{4}\right) \\
&=&\beta _{0}+\beta _{2}x^{2}+\beta _{4}x^{4}+\beta _{6}x^{6}+\beta _{8}x^{8}
\end{eqnarray*}%
with:%
\begin{eqnarray*}
\beta _{0} &=&b_{1,0}b_{2,0} \\
\beta _{2} &=&b_{1,0}b_{2,2}+b_{1,2}b_{2,0} \\
\beta _{4} &=&b_{1,0}b_{2,4}+b_{1,2}b_{2,2}+b_{2,0}b_{1,4} \\
\beta _{6} &=&b_{1,4}b_{2,2}+b_{1,2}b_{2,4} \\
\beta _{8} &=&b_{1,4}b_{2,4}
\end{eqnarray*}

So we need to compute the following integrals, which can be obtained from
the first moments of the Gaussian distribution:%
\begin{eqnarray*}
I_{0} &=&\beta _{0}\int_{-\infty }^{+\infty }\frac{1}{\sigma \sqrt{2\pi }}%
\exp \left[ -\frac{1}{2\sigma ^{2}}\left( s_{t}-\mu _{t}\right) ^{2}\right]
ds_{t}=\beta _{0} \\
I_{2t} &=&\beta _{2}\int_{-\infty }^{+\infty }\frac{1}{\sigma \sqrt{2\pi }}%
s_{t}^{2}\exp \left[ -\frac{1}{2\sigma ^{2}}\left( s_{t}-\mu _{t}\right) ^{2}%
\right] ds_{t}=\beta _{2}\left( \mu _{t}^{2}+\sigma ^{2}\right) =\beta
_{2}\left( \left( \frac{h_{1t}}{\sigma _{1}^{2}}+\frac{h_{2t}}{\sigma
_{2}^{2}}\right) ^{2}\sigma ^{4}+\sigma ^{2}\right) \\
I_{4t} &=&\beta _{4}\int_{-\infty }^{+\infty }\frac{1}{\sigma \sqrt{2\pi }}%
s_{t}^{4}\exp \left[ -\frac{1}{2\sigma ^{2}}\left( s_{t}-\mu \right) ^{2}%
\right] ds_{t}=\beta _{4}\left( \mu _{t}^{4}+6\mu _{t}^{2}\sigma
^{2}+3\sigma ^{4}\right) \\
I_{6t} &=&\beta _{6}\int_{-\infty }^{+\infty }\frac{1}{\sigma \sqrt{2\pi }}%
s_{t}^{6}\exp \left[ -\frac{1}{2\sigma ^{2}}\left( s_{t}-\mu _{t}\right) ^{2}%
\right] ds_{t}=\beta _{6}\left( \mu _{t}^{6}+15\mu _{t}^{4}\sigma ^{2}+45\mu
_{t}^{2}\sigma ^{4}+15\sigma ^{6}\right) \\
I_{8t} &=&\beta _{8}\int_{-\infty }^{+\infty }\frac{1}{\sigma \sqrt{2\pi }}%
s_{t}^{8}\exp \left[ -\frac{1}{2\sigma ^{2}}\left( s_{t}-\mu _{t}\right) ^{2}%
\right] ds_{t}=\beta _{8}\left( 
\mu
_{t}^{8}+28%
\mu
_{t}^{6}\sigma ^{2}+210%
\mu
_{t}^{4}\sigma ^{4}+420%
\mu
_{t}^{2}\sigma ^{6}+105\sigma ^{8}\right)
\end{eqnarray*}

Finally, we have that:%
\begin{eqnarray}
\Lambda _{ML} &=&\underset{\alpha ,\sigma _{1},\sigma _{2},c_{1,4},c_{2,4}}{%
\max }\prod\limits_{t=1}^{T}\frac{\sigma }{\alpha }\frac{\overline{\sigma }%
_{1}\overline{\sigma }_{2}}{\sigma _{1}\sigma _{2}}\exp \left[ -\frac{%
h_{1t}^{2}}{2\sigma _{1}^{2}}-\frac{h_{2t}^{2}}{2\sigma _{2}^{2}}+1\right]
\exp \left[ \frac{1}{2}\sigma ^{2}\left( \frac{h_{1t}}{\sigma _{1}^{2}}+%
\frac{h_{2t}}{\sigma _{2}^{2}}\right) ^{2}\right] \\
&&\times \left( I_{0}+I_{1t}+I_{2t}+I_{3t}+I_{4t}+I_{6t}+I_{8t}\right)
\end{eqnarray}

We note that when $c_{1,4}=c_{2,4}=0$, that is when the third and
fourth-order cumulant vanish, as would be the case for a Gaussian
distribution, then we have $I_{0}=1$, $I_{1}=I_{2}=I_{4}=I_{6}=I_{8t}=0,$
and we recover the maximum likelihood statistic of the Gaussian case:

\begin{equation}
\Lambda _{ML}^{G}=\underset{\alpha ,\sigma _{1},\sigma _{2}}{\max }%
\prod\limits_{t=1}^{T}\frac{\sigma }{\alpha }\frac{\overline{\sigma }_{1}%
\overline{\sigma }_{2}}{\sigma _{1}\sigma _{2}}\exp \left[ -\frac{h_{1t}^{2}%
}{2\sigma _{1}^{2}}-\frac{h_{2t}^{2}}{2\sigma _{2}^{2}}+1\right] \exp \left[ 
\frac{1}{2}\sigma ^{2}\left( \frac{h_{1t}}{\sigma _{1}^{2}}+\frac{h_{2t}}{%
\sigma _{2}^{2}}\right) ^{2}\right]
\end{equation}

In general, the presence of the additional terms implies a correction with
respect to the Gaussian case. In the Gaussian case, one obtains the
following explicit expressions for the variables involved in the
maximization of the likelihood detection statistic \cite{2003PhRvD..67h2003D}%
: 
\begin{eqnarray*}
\alpha ^{2} &=&\widehat{\alpha }^{2}\equiv \left( \overline{\alpha }%
^{2}\right) ^{+} \\
\sigma _{i}^{2} &=&\widehat{\sigma }_{i}^{2}\equiv \left( \overline{\sigma }%
_{i}^{2}-\widehat{\alpha }^{2}\right) ^{+}
\end{eqnarray*}

Here, the expression for the likelihood detection statistic is more involved
and since it is not clear whether any analytical solutions can be obtained
for the values for $\alpha ,\sigma _{1},\sigma _{2},c_{1,4},c_{2,4}$ that
would lead to the maximum in $\Lambda _{ML}$, one would need to resort to
numerical optimization procedures. Taking the log, we have that:

\begin{equation}
\log \Lambda _{ML}=\underset{\alpha ,\sigma _{1},\sigma _{2},c_{1,4},c_{2,4}}%
{\max }\log \Lambda _{ML}^{G}+\sum\limits_{t=1}^{T}\log \left(
I_{0}+I_{1t}+I_{2t}+I_{3t}+I_{4t}+I_{6t}+I_{8t}\right)  \label{logMLGC2}
\end{equation}

\subsection{Non-Gaussian Signal and Noise Distributions}

The methodology can also be extended to account for the presence of
deviations from the Gaussian assumption for both the signal and noise
distributions. To do so, we use again the Edgeworth expansion to approximate
the unknown noise distribution as $f_{s}\left( x\right) \simeq
f_{s}^{G}\left( x\right) g_{s}\left( x\right) $ and $f_{n_{i}}\left(
x\right) \simeq f_{n_{i}}^{G}\left( x\right) g_{n_{i}}\left( x\right) $ with:%
\begin{eqnarray}
f_{s}^{G}\left( x\right) &\equiv &\frac{1}{\sqrt{2\pi }\alpha }\exp \left[ -%
\frac{x^{2}}{2\alpha ^{2}}\right] \\
g_{s}\left( x\right) &\equiv &\underset{b_{0}}{\underbrace{\left( 1+\frac{%
c_{4}}{8\alpha ^{4}}-\frac{5c_{3}^{2}}{24\alpha ^{6}}\right) }}\underset{%
b_{1}}{\underbrace{-\frac{c_{3}}{2\alpha ^{4}}}}x\underset{b_{2}}{+%
\underbrace{\left( \frac{15c_{3}^{2}}{24\alpha ^{8}}-\frac{c_{4}}{4\alpha
^{6}}\right) }}x^{2}+\underset{b_{3}}{\underbrace{\frac{c_{3}}{6\alpha ^{6}}}%
}x^{3}+\underset{b_{4}}{\underbrace{\left( \frac{c_{4}}{24\alpha ^{8}}-\frac{%
5c_{3}^{2}}{24\alpha ^{10}}\right) }}x^{4}+\underset{b_{6}}{\underbrace{%
\frac{c_{3}^{2}}{72\alpha ^{6}}}}x^{6} \\
f_{n_{i}}^{G}\left( x\right) &\equiv &\frac{1}{\sqrt{2\pi }\sigma _{i}}\exp %
\left[ -\frac{x^{2}}{2\sigma _{i}^{2}}\right] \\
g_{n_{i}}\left( x\right) &\equiv &\underset{b_{i,0}}{\underbrace{\left( 1+%
\frac{c_{i,4}}{8\sigma _{i}^{4}}-\frac{5c_{i,3}^{2}}{24\sigma _{i}^{6}}%
\right) }}\underset{b_{i,1}}{\underbrace{-\frac{c_{i,3}}{2\sigma _{i}^{4}}}}x%
\underset{b_{i,2}}{+\underbrace{\left( \frac{15c_{i,3}^{2}}{24\sigma _{i}^{8}%
}-\frac{c_{i,4}}{4\sigma _{i}^{6}}\right) }}x^{2}+\underset{b_{i,3}}{%
\underbrace{\frac{c_{i,3}}{6\sigma _{i}^{6}}}}x^{3}+\underset{b_{i,4}}{%
\underbrace{\left( \frac{c_{i,4}}{24\sigma _{i}^{8}}-\frac{5c_{i,3}^{2}}{%
24\sigma _{i}^{10}}\right) }}x^{4}+\underset{b_{i,6}}{\underbrace{\frac{%
c_{i,3}^{2}}{72\sigma _{i}^{6}}}}x^{6}
\end{eqnarray}

In this context, the likelihood maximization problem becomes:%
\begin{eqnarray*}
\Lambda _{ML} &=&\underset{%
\begin{array}{c}
\alpha ,\sigma _{1},\sigma _{2},c_{3},c_{4}, \\ 
c_{1,3},c_{1,4},c_{2,3},c_{2,4}%
\end{array}%
}{\max }\prod\limits_{t=1}^{T}\frac{\overline{\sigma }_{1}\overline{\sigma }%
_{2}}{\sqrt{2\pi }\alpha \sigma _{1}\sigma _{2}}\int_{-\infty }^{+\infty
}\exp \left[ -\frac{s_{t}^{2}}{2\alpha ^{2}}-\frac{\left(
h_{1t}-s_{t}\right) ^{2}}{2\sigma _{1}^{2}}-\frac{\left( h_{2t}-s_{t}\right)
^{2}}{2\sigma _{2}^{2}}+1\right] g_{s}\left( x\right) g_{n_{1}}\left(
x\right) g_{n_{2}}\left( x\right) ds_{t} \\
&=&\underset{%
\begin{array}{c}
\alpha ,\sigma _{1},\sigma _{2},c_{3},c_{4}, \\ 
c_{1,3},c_{1,4},c_{2,3},c_{2,4}%
\end{array}%
}{\max }\prod\limits_{t=1}^{T}\frac{\sigma }{\alpha }\frac{\overline{\sigma }%
_{1}\overline{\sigma }_{2}}{\sigma _{1}\sigma _{2}}\exp \left[ -\frac{%
h_{1t}^{2}}{2\sigma _{1}^{2}}-\frac{h_{2t}^{2}}{2\sigma _{2}^{2}}+1\right]
\exp \left[ \frac{1}{2}\sigma ^{2}\left( \frac{h_{1t}}{\sigma _{1}^{2}}+%
\frac{h_{2t}}{\sigma _{2}^{2}}\right) ^{2}\right] \\
&&\times \int_{-\infty }^{+\infty }\frac{1}{\sigma \sqrt{2\pi }}\exp \left[ -%
\frac{1}{2\sigma ^{2}}\left( s_{t}-\mu _{t}\right) ^{2}\right] g_{s}\left(
x\right) g_{n_{1}}\left( x\right) g_{n_{2}}\left( x\right) ds_{t}
\end{eqnarray*}%
with:%
\begin{eqnarray*}
\sigma &=&\left( \frac{1}{\alpha ^{2}}+\frac{1}{\sigma _{1}^{2}}+\frac{1}{%
\sigma _{2}^{2}}\right) ^{-\frac{1}{2}} \\
\mu _{t} &=&\left( \frac{h_{1t}}{\sigma _{1}^{2}}+\frac{h_{2t}}{\sigma
_{2}^{2}}\right) \sigma ^{2}
\end{eqnarray*}

Focussing for simplicity of exposure on symmetric noise distribution
functions, for which we have $c_{3},c_{1,3},c_{2,3}=0$, we obtain:%
\begin{eqnarray*}
g_{s}\left( x\right) g_{n_{1}}\left( x\right) g_{n_{2}}\left( x\right)
&=&\left( b_{0}+b_{2}x^{2}+b_{4}x^{4}\right) \left(
b_{1,0}+b_{1,2}x^{2}+b_{1,4}x^{4}\right) \left(
b_{2,0}+b_{2,2}x^{2}+b_{2,4}x^{4}\right) \\
&=&\gamma _{0}+\gamma _{2}x^{2}+\gamma _{4}x^{4}+\gamma _{6}x^{6}+\gamma
_{8}x^{8}+\gamma _{8}x^{10}+\gamma _{8}x^{12}
\end{eqnarray*}%
with straightforward expressions for the $\gamma _{i}$ terms as a function
of the $b_{1,i}$, $b_{2,i}$ and $b_{i}$ coefficients.

So we need to compute the following integrals, which can be obtained from
the first moments of the Gaussian distribution:%
\begin{eqnarray*}
I_{0} &=&\gamma _{0}\int_{-\infty }^{+\infty }\frac{1}{\sigma \sqrt{2\pi }}%
\exp \left[ -\frac{1}{2\sigma ^{2}}\left( s_{t}-\mu _{t}\right) ^{2}\right]
ds_{t} \\
I_{2t} &=&\gamma _{2}\int_{-\infty }^{+\infty }\frac{1}{\sigma \sqrt{2\pi }}%
s_{t}^{2}\exp \left[ -\frac{1}{2\sigma ^{2}}\left( s_{t}-\mu _{t}\right) ^{2}%
\right] ds_{t} \\
I_{4t} &=&\gamma _{4}\int_{-\infty }^{+\infty }\frac{1}{\sigma \sqrt{2\pi }}%
s_{t}^{4}\exp \left[ -\frac{1}{2\sigma ^{2}}\left( s_{t}-\mu \right) ^{2}%
\right] ds_{t} \\
I_{6t} &=&\gamma _{6}\int_{-\infty }^{+\infty }\frac{1}{\sigma \sqrt{2\pi }}%
s_{t}^{6}\exp \left[ -\frac{1}{2\sigma ^{2}}\left( s_{t}-\mu _{t}\right) ^{2}%
\right] ds_{t} \\
I_{8t} &=&\gamma _{8}\int_{-\infty }^{+\infty }\frac{1}{\sigma \sqrt{2\pi }}%
s_{t}^{8}\exp \left[ -\frac{1}{2\sigma ^{2}}\left( s_{t}-\mu _{t}\right) ^{2}%
\right] ds_{t} \\
I_{10t} &=&\gamma _{10}\int_{-\infty }^{+\infty }\frac{1}{\sigma \sqrt{2\pi }%
}s_{t}^{10}\exp \left[ -\frac{1}{2\sigma ^{2}}\left( s_{t}-\mu _{t}\right)
^{2}\right] ds_{t} \\
I_{12t} &=&\gamma _{12}\int_{-\infty }^{+\infty }\frac{1}{\sigma \sqrt{2\pi }%
}s_{t}^{8}\exp \left[ -\frac{1}{2\sigma ^{2}}\left( s_{t}-\mu _{t}\right)
^{2}\right] ds_{t}
\end{eqnarray*}%
\newline

Again, we need to compute higher-order moments of the Gaussian distribution,
which is given by the following formula for a normally distributed variable $%
X$ with mean $\mu $ and variance:%
\begin{equation}
\mathbb{E}\left( X^{n}\right) =\int_{-\infty }^{+\infty }\frac{1}{\sigma 
\sqrt{2\pi }}x^{n}\exp \left[ -\frac{1}{2\sigma ^{2}}\left( n-\mu \right)
^{2}\right] dx=\sum\limits_{j=0}^{\left[ \frac{n}{2}\right] }\binom{n}{2j}%
\left( 2j-1\right) !!\sigma ^{2j}\mu ^{n-2j}
\end{equation}%
where $n!!$ denotes the double factorial operator $n!!=\prod%
\limits_{i=0}^{k}(n-2i)=n\left( n-2\right) \left( n-4\right) ...$ with $k=%
\left[ \frac{n}{2}\right] $.

For example, we have that:%
\begin{equation}
I_{10t}=\gamma _{10}\left( 
\mu
_{t}^{10}+45%
\mu
_{t}^{8}\sigma ^{2}+630%
\mu
_{t}^{6}\sigma ^{4}+3150%
\mu
_{t}^{4}\sigma ^{6}+4725%
\mu
_{t}^{2}\sigma ^{8}+945\sigma ^{10}\right)
\end{equation}

Finally, we have that:%
\begin{eqnarray}
\Lambda _{ML} &=&\underset{%
\begin{array}{c}
\alpha ,\sigma _{1},\sigma _{2},c_{3},c_{4}, \\ 
c_{1,3},c_{1,4},c_{2,3},c_{2,4}%
\end{array}%
}{\max }\prod\limits_{t=1}^{T}\frac{\sigma }{\alpha }\frac{\overline{\sigma }%
_{1}\overline{\sigma }_{2}}{\sigma _{1}\sigma _{2}}\exp \left[ -\frac{%
h_{1t}^{2}}{2\sigma _{1}^{2}}-\frac{h_{2t}^{2}}{2\sigma _{2}^{2}}+1\right]
\exp \left[ \frac{1}{2}\sigma ^{2}\left( \frac{h_{1t}}{\sigma _{1}^{2}}+%
\frac{h_{2t}}{\sigma _{2}^{2}}\right) ^{2}\right]  \notag \\
&&\times \left(
I_{0}+I_{1t}+I_{2t}+I_{3t}+I_{4t}+I_{6t}+I_{8t}+I_{10t}+I_{12t}\right)
\end{eqnarray}

We note that when $c_{i,3}=c_{i,4}=0$, that is when the third and
fourth-order cumulant vanish for the noise distribution, we then recover the
maximum likelihood statistic from \cite{2014PhRvD..89l4009M}.

\section{Conclusions and Extensions}

This paper analyzes the comparative efficiency of the standard CC detection
statistic versus an alternative detection statistic obtained by
cross-correlating squared measurements in situations involving non-Gaussian
noise (and signal) distributions and heterogeneous detector sensitivities.
We find that differences in detector sensitivities have a large impact on
the efficiency of the CC detection statistic, which is dominated by the
alternative statistic when these differences reach one order of magnitude.
This effect is smaller in case of fat-tailed noise distributions, but it is
magnified in case noise distributions have skewness parameters of opposite
signs. On the other hand, higher-order cumulants of the signal distribution
do not have a material impact on the relative efficiency of the two
detection statistics in realistic situations where the signal is expected to
be small compared to the noise. Since our methodology requires the
estimation of higher-order moments/cumulants of the noise distribution, we
extend the maximum likelihood estimator to the case of non-Gaussian signal
and noise distributions and manage to recover analytical expressions for the
log-likelihood function in case these distributions can be approximated by
Edgeworth-type expansions.

Our methodology can be extended in a number of directions. We may first
consider a setting involving a correlated noise component, typically
regarded as environmental noise, in addition to the specific instrumental
noise. On a different note, we have considered so far colocated and
coincident detectors, an assumption which would hold in the case of Einstein
Telescope. On the other hand, our framework should be extended to apply to a
network of separated detectors such as Advanced LIGO-Virgo detectors, or
joint observations by Advanced LIGO and Einstein Telescope. This extension
is important because these are precisely the types of situations where
differences in sensitivities are expected to be most substantial.

\begin{acknowledgments}
We are thankful to an anonymous referee whose comments have significantly improved the quality of this work.
\end{acknowledgments}
\bibliographystyle{plain}
\bibliography{biblio}

\end{document}